\documentclass[12pt]{article}

\usepackage{amsmath, amsfonts, amssymb, mathrsfs, siunitx}
\usepackage{dcolumn}
\usepackage[authoryear]{natbib}
\usepackage{float}
\usepackage{indentfirst}
\usepackage{subcaption}
\usepackage{amsmath,bbm,amssymb}

\DeclareMathOperator*{\argmin}{arg\,min}
\usepackage{amsthm,amscd}
\usepackage{graphicx, graphics}
\usepackage{comment}
\usepackage{longtable, lscape}
\usepackage{rotating}
\usepackage[page]{appendix}
\usepackage{multirow}
\usepackage{color}

\usepackage{cleveref}

\newtheorem{theorem}{Theorem}
\newtheorem{lemma}{Lemma}

\newtheorem{algorithm}{Algorithm}

\renewcommand{\baselinestretch}{1.5}
\newcommand{\bm}[1]{\mbox{\boldmath$ #1 $\unboldmath}}

\usepackage{algorithm}
\usepackage{algpseudocode}

\makeatletter
\def\BState{\State\hskip-\ALG@thistlm}
\makeatother

\makeatletter
\renewcommand{\fnum@algorithm}{\fname@algorithm}
\makeatother

\begin{document}

\def\spacingset#1{\renewcommand{\baselinestretch}%
{#1}\small\normalsize} \spacingset{1.25}

\title{\bf Neighborhood VAR: Efficient estimation of multivariate timeseries with neighborhood information}
\author{Zhihao Hu\textsuperscript{1}, Shyam Ranganathan\textsuperscript{2}
	, Yang Shao\textsuperscript{3}, Xinwei Deng\textsuperscript{1}}
   
    \date{{\small \textsuperscript{1}Department of Statistics, Virginia Tech, Blacksburg, VA 24061, USA.\\
    \textsuperscript{2}School of Mathematical and Statistical Sciences, Clemson University, Clemson, SC 29634, USA.\\
    \textsuperscript{3}Department of Geography, Virginia Tech, Blacksburg, VA 24061, USA.}}
\maketitle

\begin{abstract}
	In data science, vector autoregression (VAR) models are popular in modeling multivariate time series in the environmental sciences and other applications. However, these models are computationally complex with the number of parameters scaling quadratically with the number of time series.
	In this work, we propose a so-called neighborhood vector autoregression (NVAR) model to efficiently analyze large-dimensional multivariate time series.
	We assume that the time series have underlying neighborhood relationships, e.g., spatial or network, among them based on the inherent setting of the problem. When this neighborhood information is available or can be summarized using a distance matrix, we demonstrate that our proposed NVAR method provides a computationally efficient and theoretically sound estimation of model parameters. The performance of the proposed method is compared with other existing approaches in both simulation studies and a real application of stream nitrogen study.\\
\end{abstract}





\section{Introduction}

Modeling multivariate time series has attracted great attention from different areas such as environmental sciences (e.g. \citet{Davis,Cavalcante2017}), network sciences (e.g. \citet{sensorVAR,ValdesSosa2005}), gene expression (e.g. \citet{Opgen-Rhein2007, network}), and economics (e.g. \citet{econometrics1,econometrics2,econometrics3,StructuredVAR}). With the increasing sophistication in data availability and methodology, recent developments in advanced manufacturing such as \citet{falcon, hsu2020, ghahramani2020ai} have also started focusing on multivariate time series analysis. 

In a number of these multivariate timeseries problems, the underlying source of the timeseries data exhibits dependencies either in the form of a spatial field, which can then be handled by spatiotemporal methods or Gaussian processes e.g., \citet{Wikle}, or neighborhood information, where ``nearby'' timeseries are more correlated with the focal timeseries than ``farther'' ones, e.g., \citet{BandedVAR}. This notion of ``neighborhood'' is formalized in this paper based on the context of the problem, i.e., spatial timeseries or networked timeseries etc. using a distance matrix and, under the assumption that this distance matrix is known, this paper develops a new methodology to efficiently model multivariate timeseries. In cases where the distance matrix is not given, this paper provides heuristics to compute it based on the structural relationships of the timeseries for the problem at hand. The theoretical justification for the estimation method, a simulation study and an interesting illustration to an important management problem in water systems control using a stream nitrogen study are also demonstrated in this paper.

When the number of time series is large, the number of parameters in the conventional vector autoregression model (VAR) increases dramatically, posing great challenges to performing proper estimation and inference on these time series data. Recent work to address this issue in the context of high-dimensional data has resulted in the notion of sparsity for efficient estimation of the VAR model, including the use of lasso penalization (e.g. \citet{Hsu2008,Arnold2007,Lozano2009,
	largeVAR}), the group lasso penalization (e.g. \citet{Haufe, Shojaie2010, Basu2015,Bolstad2011}), Dantzig-type penalization (e.g. \citet{Qiu2015,Han2015}), adaptive lasso penalization (e.g. \citet{Ren2010, Kock2014}), graph regularization (e.g. \citet{Jiang2015}), among many others. Such strategies typically adopt regularization to reduce the the number of parameters in the estimated model, achieving model parsimony.

In many applications with multiple time series, e.g., modeling the nitrogen of multiple streams that is considered here, there is a natural dependence structure among time series collected from neighboring locations. The spatial nature of the problem implies that two time series from nearby locations are more likely to correlate with each other than those from far-away locations. Without loss of generality, let us assume that each time series is collected from a sensor at a location. Here we use the metaphor of a ``sensor'' though the so-called sensor may not have a physical meaning, and just represents the location where the time series originated from. The term ``location'' is also not restricted to physical location, and instead means that the multiple timeseries are assumed to have a correlation with each other that can be measured in terms of their distances under some distance measure in a metric space. For instance, for time series data over networks, one could consider an embedding of the network in a latent space (e.g. \citet{hoff2002latent}) and measure distances between nodes based on the distances in the latent space. This is different from the typical spatial statistics or spatio-temporal statistics problem, where there is an underlying spatial process that results in spatial observations at different locations.

In the work of \citet{BandedVAR}, the time series are assumed to be located on a one-dimensional lattice at equally spaced intervals.
Their so-called banded VAR (BVAR) method has a clear interpretation and can effectively reduce model complexity while outperforming penalization-based algorithms like the LASSO.
But the assumption of a sequential ordering of time series is too restrictive for general applications and
it would be more reasonable to consider a more generalized notion of location on a vector space rather than on a one-dimensional lattice.

In this article, we propose the Neighborhood Vector AutoRegression model (NVAR), which extends the notion of ``band'' in \citet{BandedVAR} to the notion of a ``neighborhood''.
Having a more general modeling assumption, the proposed method maintains efficient parameter estimation with clear model interpretation. The proposed method also guarantees the convergence rate of the estimated coefficient matrix. In particular, we show that the asymptotic properties proved in the \citet{BandedVAR} paper hold for our proposed method.
In the case study of modeling the nitrogen content of multiple streams, the proposed NVAR method takes advantage of the notion of neighborhood to model the water quality data from multiple streams in a joint manner. In the United States, excess nutrients are one of the most important causes of impairment for rivers and streams (see \citet{EPA2000}).
Increasing rates of nutrient supply fuels accelerating primary production or eutrophication, which leads to discoloration of affected waters (see \citet{Paerl2001}).
The proposed method provides a useful tool to understand how the water quality changes over time and also the interactions of water systems at different locations. In comparison with several existing approaches, the case study shows the merits of the proposed method with a reasonable computational cost.

The remainder of the article is organized as follows.
Section 2 describes the neighborhood VAR model and the algorithm.
Section 3 provides some theoretical results for the proposed method.
Section 4 discusses simulation results and we conclude this work with a practical application in Section 5.

\vspace{-0.25cm}
\section{The Proposed Model}
Denote $\mathbf{y}(t)=[y_1(t),y_2(t),\ldots,y_p(t)]^T$ as the $p$ dependent time series, where $y_i(t) \in \mathbb{R}$ and $t=1,2,\ldots, n$.
We assume that the time series $y_1(t),\ldots, y_p(t)$ are collected from ``sensors'' located at $s_1,\ldots,s_p$ with $s_i \in (M,d)$, where $M$ is a metric space with distance measure defined by $d$, say $(\mathbb{R}^m,d)$.
Here the $d$ is a pre-defined distance measure with distances $d(s_i,s_j)$ for $s_i, s_j \in M$. Hereafter we will abbreviate $d(s_i,s_j)$ as $d(i,j)$ for notation convenience.
Since the sensors have a distance measure, we can define ``$d_0$-neighborhood'' of the $i^{th}$ time series as $\mathcal{N}_{i}^{d_0} = \{j:d(i,j)\le d_0\}$ for some $d_0 \in R$, and is a representation of the set of all time series that have an influence on the $i^{th}$ time series at this particular distance level $d_0$.
Thus we consider the NVAR($q$) model in a form similar to the classical VAR(q) model as
\begin{align}\label{eq: AR_model}
\mathbf{y}(t) = A_1\mathbf{y}(t-1) + A_2\mathbf{y}(t-2) + \ldots + A_q\mathbf{y}(t-q) + \mathbf{e}(t), t=1,2,\ldots, n,
\end{align}
with  $A_s(i,j) = 0, s=1,\ldots,q, j \notin \mathcal{N}_{i}^{d_0}$ for some ``$d_0$-neighborhood'' of $i$, $\mathcal{N}_{i}^{d_0}$. Here $q$ is the lag order for the autoregressive process.  The coefficient matrices $A_1,\ldots, A_q$ are $p\times p$ matrices that represent the dependence between the different time series. We do not use bold typeface for $A$ to enhance readability. The $\mathbf{e}(t)$ is the serially uncorrelated noise with $E(\mathbf{e})=0$ and $var(\mathbf{e})=\Sigma_e$. Note that we need to assume conditions on the coefficient matrices for the time series to be stationary, and we make the assumption that $|I-A_1z-...-A_qz^q|\ne 0$ for any $|z| \le 1$.
Hereafter, we will drop the time index $t$ where possible to enhance readability.

Note that our proposed neighborhood idea generalizes the notion of bandwidth in the banded VAR model and has parallels to work by \citet{Besag1974} among others.
The proposed NVAR model considers the time series to be homogeneous in the sense that the same $d_0$ is sufficient to characterize the neighbors' influence for every time series.
It can also be extended to the case with different values of $d_0$ being estimated for each individual time series $y_i$ but we do not consider that case in this paper.
When there is no direct mapping of time series locations to a metric space, we can potentially embed these time series in a latent space, and obtain distances from this embedded space. This adds another layer of uncertainty due to estimation of the distance matrix itself. However, in the rest of this paper, we will not deal with these technical complications and we assume that the distances are well-specified and given by the symmetric $p\times p$ matrix $\mathbf{D}$. Note that time-varying distances are allowed in this algorithm as the sensors may be allowed to move in time. This can be included in our algorithm by specifying a different matrix $\mathbf{D(t)}$ at each time instant. For easy comprehension though, in this paper, we assume that the coefficient matrices are time-invariant.


\vspace{-0.5cm}
\subsection{Construction of Neighborhood}
For every time series $y_i$, we define a $d$-neighborhood as the set of all time series that are at most $d$ distance away from it. This is denoted by
\[
\mathcal{N}_i^d = \{j:d(i,j)\le d\}.
\]
The neighborhood VAR model assumes that $y_i$ depends only on its $d_0-neighborhood$ for some value of $d = d_0$. This is a generalization of bandwidth in the banded VAR and allows us to handle more complex time series. Specifically, if we assume that the time series reside on a $1$-D lattice at locations $1,2,\ldots,p$, and distance is measured as $d(i,j) = |i-j|$, we get back the banded VAR formulation with bandwidth $=d_0$.

Our definition of neighborhood VAR model assumes that the elements of the coefficient matrix are non-zero only at locations that are within the $d_0-neighborhood$ of each of the time series $i$. That is,
\[A_r(i,j) = 0,\ \text{if}\ j \notin \mathcal{N}_{i}^{d_0},\]
where $A_r(i,j)$ represents the $(i,j)^{th}$ element of the coefficient matrix $A_r$ corresponding to a lag of $r$. Thus the maximum number of non-zero elements in row $i$ of the coefficient matrices is given by $\tau_{i} = |\mathcal{N}_{i}^{d_0}|$. The sparsity assumption implies that $\tau_{i} \ll p$.


Below are a few illustrative cases on constructing the neighborhood.

{\bf Banded Structures:} The banded matrix structure of $A_{r}$ discussed in the banded VAR literature is obtained if the time series are assumed to be present in locations that are arranged along a line segment such that $s_1<s_2<\ldots<s_p$ and $d(s_i,s_j)=|i-j|$.


{\bf Block-banded Structures:} A block banded structure is obtained if the locations of the time series correspond to locations arranged in 2-D space (e.g., pixels in an image matrix) with equal distances between neighboring locations and $d(s_i,s_j)=|i-j| \mod{\sqrt{p}}$.
Here the locations are in a $\sqrt{p} \times \sqrt{p}$ lattice and the $s_{i}$ are obtained via vectorizing the lattices into a p-dimensional set of locations in a row-by-row fashion.
Note that this is equivalent to consider each location to be affected by geographically close locations as measured using a city-block distance metric. Other formulations and distance metrics lead to different structures on 2-D data.
%

{\bf Neighborhood Structure under Spatial Data:} In spatial-temporal data, the Euclidean distance between the sensor locations can be used for the distance matrix of sensors.
The sparsity structure of the coefficient matrices will  depend on the actual distances between the sensors.
Note that it is important, in this case, to normalize the distances to avoid identifiability problems in terms of the scale, and also to preserve the spatial meaning of neighborhood in terms of the actual problem.
A good scaling constant that can be used will approximate the spatial scale to a lattice by scaling distances as $\frac{(N/2)}{(d_{max})^m}$,
where $N$ is the number of timeseries, $m$ is the dimension of the space ($1$, $2$ or $3$ for typical spatial problems), and $d_{max}$ is the maximum distance among all pairs of sensors.

{\bf Neighborhood Structure under Network Data:} In a network application, the time series are often from sensors that are connected to each other under a network.
The distance matrix can be specified by the adjacency matrix of the network, with the length of the shortest path between two nodes giving the distance between the two nodes.
For a more general formulation, we can embed the network in a latent space and compute the distance metric based on distances on the latent space.

\subsection{Parameter Estimation}
Given a particular value of $d_0$, and subsequently $\mathcal{N}_{i}^{d_0}$ (which can be computed directly since $\mathbf{D}$ is assumed to be known), 
the estimation of the NVAR model can be conducted using the ordinary least squares (OLS) estimation of the coefficients corresponding to each time series.
For instance, if $A_{i}$ is the set of all coefficients to be estimated for the $i^{th}$ time series, we can obtain $A_i$ from $\mathcal{N}_{i}^{d_0}$ and the lag order $q$ by selecting the appropriate elements from the VAR coefficient matrices. In the simple case of VAR(1), where the only coefficient matrix is $A_1$, the OLS equation for the $i^{th}$ time series is simply
\[
\hat{y}_i(t) = \sum_{j \in \mathcal{N}_{i}^{d_0}, r \in \{1,\ldots,q\}} \hat{A}_1(i,j) y_j(t-r),
\]
where $\hat{A}_1(i,j)$ is the $(i,j)^{th}$ element in the estimated coefficient matrix $\hat{A}_1$, and hence the estimates for the coefficient matrices are obtained in a straightforward manner, with all elements of the matrix outside the $d_0$-neighborhood set to $0$.

For NVAR(q) model, we can also use the ordinary least squares (OLS) estimation to estimate parameters separately in each time series variables with respect to its $d_0-neighborhood$.
Let us denote 
\begin{align*}
&\textbf{y}_i=\begin{pmatrix}
y_i(n) \\
y_i(n - 1) \\
\vdots \\
y_i(q + 1)
\end{pmatrix}
\end{align*}
\begin{align*}
\textbf{X}_i=\begin{pmatrix}
\{\textbf{y}_{j}(n - 1)\}^T\big|_{j \in \mathcal{N}^{d_0}_i} & \{\textbf{y}_{j}(n - 2)\}^T\big|_{j \in \mathcal{N}^{d_0}_i} & \cdots & \{\textbf{y}_{j}(n - q)\}^T\big|_{j \in \mathcal{N}^{d_0}_i} \\
\{\textbf{y}_{j}(n - 2)\}^T\big|_{j \in \mathcal{N}^{d_0}_i} & \{\textbf{y}_{j}(n - 3)\}^T\big|_{j \in \mathcal{N}^{d_0}_i} & \cdots & \{\textbf{y}_{j}(n - q - 1)\}^T\big|_{j \in \mathcal{N}^{d_0}_i} \\
\vdots & \vdots & \vdots & \vdots \\
\{\textbf{y}_{j}(q)\}^T\big|_{j \in \mathcal{N}^{d_0}_i} & \{\textbf{y}_{j}(q - 1)\}^T\big|_{j \in \mathcal{N}^{d_0}_i} & \cdots & \{\textbf{y}_{j}(1)\}^T\big|_{j \in \mathcal{N}^{d_0}_i} \\
\end{pmatrix}_{(n - q) \times q |\mathcal{N}^{d_0}_i|},
\end{align*}
where $\{\textbf{y}_{j}(t)\}\big|_{j \in \mathcal{N}^{d_0}_i} = \begin{pmatrix}
y_{j_1}(t), & y_{j_2}(t), & \dots, & y_{j_{|\mathcal{N}^{d_0}_i|}}(t)
\end{pmatrix}^T,\ j_1, j_2, \dots, j_{|\mathcal{N}^{d_0}_i|} \in \mathcal{N}^{d_0}_i$ is a column vector, 
and $|\mathcal{N}^{d_0}_i| \text{ is the size of } \mathcal{N}^{d_0}_i.$
Then we denote the coefficients matrix as
\begin{align*}
&\bm B=\begin{pmatrix}
\bm\beta_{1} & \bm\beta_{2} & \cdots & \bm\beta_{p}
\end{pmatrix}_{p \times q |\mathcal{N}^{d_0}_i|},
\end{align*}
where $\bm\beta_{i}=\begin{pmatrix}
\{A_1(i,j)\}\big|_{j \in \mathcal{N}^{d_0}_i} \\
\{A_2(i,j)\}\big|_{j \in \mathcal{N}^{d_0}_i} \\
\vdots \\
\{A_q(i,j)\}\big|_{j \in \mathcal{N}^{d_0}_i}
\end{pmatrix}$, and 
$\{A_q(i,j)\}\big|_{j \in \mathcal{N}^{d_0}_i}=\begin{pmatrix}
A_q(i,j_1) \\
A_q(i,j_2) \\
\vdots \\
A_q(i,j_{|\mathcal{N}^{d_0}_i|})
\end{pmatrix}, \ j_1, j_2, \dots, j_{|\mathcal{N}^{d_0}_i|} \in \mathcal{N}^{d_0}_i$ is a column vector. 
The coefficient matrix $\bm B$ are estimated by minimizing the least-squares objective function as
\begin{align*}
\text{min}_{\bm B} \sum_{i = 1}^{p} (\textbf{y}_i - \textbf{X}_i \bm\beta_{i})^T (\textbf{y}_i - \textbf{X}_i \bm\beta_{i}).
\end{align*}
It is easy to see that the optimization can be separated into $p$ independent OLS estimation,
\begin{align*}
\text{min}_{\bm \beta_i} (\textbf{y}_i - \textbf{X}_i \bm\beta_{i})^T (\textbf{y}_i - \textbf{X}_i \bm\beta_{i}),\ i = 1, 2, \dots, p.
\end{align*}
Thus we can have
\begin{align}
&\hat{\bm\beta}_{i} = (\textbf{X}_j^T \textbf{X}_i)^{-1} \textbf{X}_i^T \textbf{y}_i,\ i = 1, 2, \dots, p.
\end{align}
With the estimates $\hat{\bm \beta}_{1}, \ldots,  \hat{\bm \beta}_{1}$, the estimated model can then be expressed as
\begin{align}
\hat{y_i}(t) = \sum_{j \in \mathcal{N}^d_i} \hat{A}_1(i,j)y_j(t-1) + \ldots \hat{A}_q(i,j)y_j(t-q),\ t=2,\ldots,n.
\end{align}

To choose the optimal value of $d_0$, we use the Bayesian information criterion (BIC) by computing
\begin{align}
BIC(d,i) = \log(RSS(d,i))+ \frac{1}{n}q\tau_iC_n\log(p\vee n),
\end{align}
for every time series $i$ and for $d=1,2,\ldots$. The value of $d$ that minimizes this quantity is the optimal value for time series $i$, i.e., $d_0(i)$.
Since we assume the $p$ time series are homogeneous, we will consider the estimate of $d$ for all time series as the maximum of all the estimated optimal neighborhood distances.
That is,
\begin{align}
\hat{d_0} = \max_{1\le i\le p} \{ \argmin_{d} {BIC(d,i)} \} 
\end{align}
In the next section, we will show that  our estimation algorithm for $d_{0}$ will lead to optimal estimation of the neighborhood distance.
Note that our estimate may result in a number of predictors being given as relevant for any time series depending on the density of arrangement in space, and hence an optional step may be used to compute a BIC among these predictors (or sany other variable selection procedure) to further reduce the number of predictors. The estimation of lag order $q$ can also be wrapped into this same algorithm by searching over a grid of $d$ and $q$ values and choosing the value of $\hat{d_0}$ and $\hat{q}$ that maximize the BIC criterion (see \citet{BandedVAR} for a similar idea).

\begin{algorithm}[H]
	\caption{Neighborhood VAR Estimation}\label{NVAR}
	\begin{algorithmic}[]\\
		\textbf{Input:} time series $y_1(t),\ldots y_p(t)$, lag order $q$, and distance matrix $\mathbf{D}$ \newline
		\textbf{Output:} Coefficient matrices: $\hat{A}_1, \ldots, \hat{A}_q$
		\For {$d\ in\ 1:d_{max}$}
		\For {$i \ in\ 1:p$}
		\State {Find the d-neighborhood $\mathcal{N}^d_i$ of the $i^{th}$ time series}
		\State {Perform regression for the $i^{th}$ time series on $\mathcal{N}^d_i$ and compute coefficients ${\beta ^{d,i}}$}
		\State {Compute the marginal BIC as $BIC(d,i) = \log(RSS(d,i))+ \frac{1}{n}q\tau_iC_n\log(p\vee n)$
			, $\tau_i$ - the number of non-zero elements in row $i$ of the coefficient matrices}
		\EndFor
		\EndFor
		\State {Find $\hat{d} = \max_{1\le i\le p} \{ \argmin_{1\le d\le d_{max}} {BIC(d,i)} \}$}
	\end{algorithmic}
	\label{alg_1}
\end{algorithm}

\section{Theoretical Properties of Estimation Consistency}
In this section, we show that, under appropriate regularity conditions, the proposed NVAR method can consistently recover the appropriate level of sparsity, in terms of the optimal neighborhood distance.
In addition, we establish the convergence rate of the estimated coefficient matrix to the true coefficient matrix.
The regularity conditions and the theorems are presented here, while the proofs are in the Appendix.

First we will formulate the NVAR(q) model of order $q$ into a NVAR(1) mode of order 1 as follows.
\begin{align*}
\tilde{\textbf{y}}(t)=\tilde{A}\tilde{\textbf{y}}(t-1)+\tilde{\textbf{e}}(t),
\end{align*}
where
\begin{align}
\tilde{\textbf{y}}(t)=\begin{pmatrix}
\textbf{y}(t) \\
\textbf{y}(t-1) \\
\vdots \\
\textbf{y}(t-q+1)
\end{pmatrix},
\tilde{A}=\begin{pmatrix}
A_1 & A_2 & \cdots & A_{q-1} & A_q \\
I_p & 0_p & \cdots & 0_p & 0_p \\
\vdots & \vdots & \vdots & \vdots & \vdots \\
0_p & 0_p & \cdots & I_p & 0_p\\
\end{pmatrix},
\tilde{\textbf{e}}(t)=\begin{pmatrix}
\textbf{e}(t) \\
0_{p\times 1} \\
\vdots \\
0_{p\times 1}
\end{pmatrix}.
\label{eq: VAR_model}
\end{align}
Such a reformulation provides a good framework for investigating the theoretical properties. Next we need several regularity conditions that are stated as follows.
Note that these regularity conditions are similar to those imposed the banded VAR approach (see \citet{BandedVAR}).
\begin{itemize}
	\item \textit{Condition} 1. For $\tilde{A}$ defined in \eqref{eq: VAR_model}, $||\tilde{A}||_2 \leq C$ and $||\tilde{A}^{j_0}||_2 \leq \delta^{j_0}$, where $C > 0$, $\delta \in (0,1)$ and $j_0 \geq 1$ are constants free of $n$ and $p$, and $j_0$ is an integer.
	\item \textit{Condition} $1'$. For $\tilde{A}$ defined in \eqref{eq: VAR_model}, $||\tilde{A}^{j_0}||_2 \leq \delta^{j_0}$, $||\tilde{A}||_{\infty} \leq C$ and $||\tilde{A}^{j_0}||_{\infty} \leq \delta^{j_0}$, where $C > 0$, $\delta \in (0,1)$ and $j_0 \geq 1$ are constants free of $n$ and $p$, and $j_0$ is an integer.
	\item \textit{Condition} 2. Let $a_{ij}^{(l)}$ be the $(i,j)$-th element of $A_l$. For each $i=1,...,p$, at least one $j \in \mathcal{N}_{i}^{d_0}$, $\{C_n \tau_{i}^{d_0} n^{-1} \text{log}(p \vee n)\}^{1/2} \ll |a_{ij}^{(l)}|$ for some $1\le l \le q$, where $C_n \rightarrow \infty$ as $n \rightarrow \infty$.
	\item \textit{Condition} 3. The minimal eigenvalue $\lambda_{\text{min}}\{\text{cov}(\textbf{y}(t))\} \ge \kappa_1$ and $\text{max}_{1\le i/le p} |\sigma_{ii}|\le \kappa_2$ for some positive constants $\kappa_1$ and $\kappa_2$ free of $p$, where $\sigma_{ii}$ is the $i$-th diagonal element of $\text{cov}(\textbf{y}(t))$, and $\lambda_{\text{min}}(\cdot)$ denotes the minimum eigenvalue.
	\item \textit{Condition} 4. The serial noise $\{\textbf{e}(t): t=1, 2, ...,n\}$ is independent and identically distributed with zero mean and covariance $\Sigma_{e}$. Furthermore, one of the two assertions below holds:
	\begin{itemize}
		\item (i) $\text{max}_{1\le i\le p}E(|\textbf{e}_i(t)|^{2q})\le C$ and $p=O(n^\beta)$, where $q>2$, $\beta \in (0,(q-2)/4)$ and $C>0$ are some constants free of $n$ and $p$;
		\item (ii) $\text{max}_{1\le i\le p}E\{\text{exp}(\lambda_0|\textbf{e}_i(t)|^{2\alpha})\}\le C$ and $\text{log}p=o(n^{\alpha/(2-\alpha)})$, where $\lambda_0>0$, $\alpha \in (0,1]$ and $C>0$ are some constants free of $n$ and $p$.
	\end{itemize}
\end{itemize}
Briefly, Condition $3$ ensures that the covariance matrix is positive definite. Condition $4(i)$ ensures strict stationarity of the process when the $e_i(t)$ are independent and identically distributed, while Condition $4(ii)$ is an identifiability condition for the minimum value among the non-zero coefficients.

Now we can show that the estimated size of neighborhood to be consistent.
\begin{theorem}
	\label{thm1}
	Assume that Conditions 1-4 hold for the the proposed neighborhood VAR. 
	Then the estimated size of the neighborhood is consistent, i.e., $Pr(\hat{d}=d_0)\rightarrow 1\ as\ n \rightarrow \infty.$
\end{theorem}
This theorem shows that the algorithm to select the optimum neighborhood distance converges to the true neighborhood distance, if it exists, as the length of the time series grows. Moreover, we can also establish the convergence error bounds between the estimated coefficient matrix $\hat{A}_{r}$ and the true coefficient matrix $A_{r}$, when using this optimum neighborhood distance in computing the estimated coefficient matrix. Using either the Frobenius norm or the $L_2$ norm, we can show bounds on the norm of the error matrix defined as the difference between the estimated coefficient matrix and the true coefficient matrix. The following theorem shows that, for each of the $q$ coefficient matrices, the error norm is bounded and goes to $0$ as the length of the time series grows with constant $p$.
\begin{theorem}
	\label{thm2}
	Assume that Conditions 1-4 hold for the proposed neighborhood VAR. Then as $n\rightarrow \infty$, we have the following error bounds of the estimated coefficient matrix as
	\begin{align*}
	||\hat{A}_j-A_j||_F & = O_p\Big\{\frac{p}{n}^{1/2}\Big\}, \\
	||\hat{A}_j-A_j||_2 & = O_p\Big\{\frac{\log p}{n}^{1/2}\Big\},
	\end{align*}
	which hold for $j=1,...,q$.
\end{theorem}
From this theorem, one can infer that the estimated coefficient matrix is accurate even when $p$ grows along with $n$ as long as the number of time series grows at a particular rate, but not as fast as $n$ as seen from the theorem. Further details are available in the Appendix.

\section{Simulation Study}
\begin{figure}[H]
	\centering
	\includegraphics[scale=0.35]{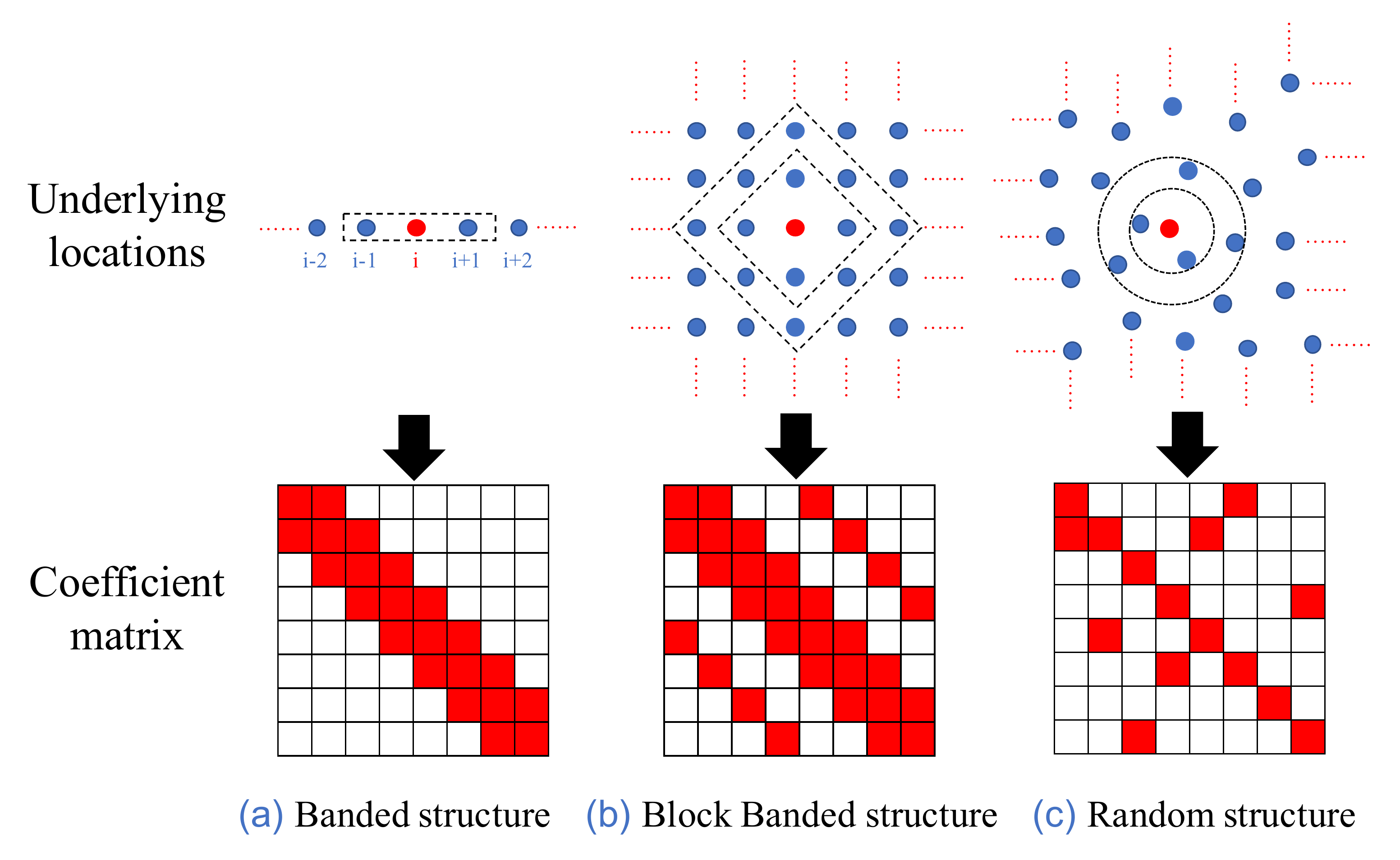}
	\vspace{-0.2cm}
	\caption{\label{fig: three-cases}Three simulation cases: banded, block-banded and random structures}
\end{figure}

In this section, we evaluate the performance of the proposed method in comparison with some existing methods.
Three different simulation cases are conducted with data generated from the NVAR(1) in \eqref{eq: AR_model}. Figure \ref{fig: three-cases} illustrates the three simulation cases.
In all these cases, we consider different values of bandwidth $d_0 = 1,2,3,4$ and different number of time series $p=100,196,400,784$.
We also assume that the distance matrix is known and that the noise process is $N(0,\sigma_{e}^{2}I_p)$ with $\sigma_{e} = 0.01, 1$.
The sample size $n$ is fixed at $n= 200$.
For each case, we create $500$ repetitions for the simulation study.

{\it Case 1: 1-D lattice structure.}
We generate random coefficient matrices ${A}$ with $A_{ij} = 0, |i-j| \ge d_0$.
The non-zero elements of the matrix are chosen randomly from $[-1,1]$ and to ensure stationarity, we force $\|A\| < 1$ by rescaling the matrix to have a random norm value between $[0.3,0.9]$ using the operation $(A/\|A\|) \times u$, where $u \sim U[0.3,0.9]$.

{\it Case 2: 2-D lattice structure.}  We generate random coefficient matrices  ${A}$ with the 2-D lattice structure shown in the previous section. From the spatial perspective, for any point $[i_1,j_1]$ the surrounding points $i_2,j_2$ with $|i_1-i_2| + |j_1-j_2| < d_0$ are the only ones that are non-zero. When we vectorize the spatial matrix to obtain the coefficient matrix $A_{r}$, this results in a block-banded structure where, for the $i^{th}$ row in $A_{r}$, the non-zero elements are the matrix positions with $|i-j| \le d_0 \pm t\sqrt{p}, t =0,1,2,\ldots \sqrt{p}, 1\le j \le p$. This means, as we increase neighborhood distance in steps, multiple time series are included, proportional to $O(d^2)$ rather than to $O(d)$ as in the 1-D lattice case. We ensure stationarity of the timeseries process as before.

{\it Case 3: 2-D spatial structure.} We generate a random spatial point process and place the ``sensors'' in these locations and then generate data based on an NVAR(1) process, where each time series depends only on its neighbors in space. To ensure a fair comparison, we maintain the density of the sensors in any small region to be similar to that of the 2-D lattice on average by suitable scaling. In this case, the number of neighbors is random and as we increase $d_0$ in steps corresponding to the lattice case, multiple time series are included in the neighborhood. For instance, we first generate a spatial point process in $[0,1] \times [0,1]$. We scale this unit square appropriately based on the particular instances of the point processes so that the average number of nearest neighbors for all points is roughly $4$. Note that, as the number of time series increases, i.e., as $p$ increases in the simulation setting, to maintain the same average number of neighbors, we need to scale the unit square differently.

The proposed method is compared with the banded VAR method and the LASSO method.
In the LASSO method, we use the Lasso regression to estimate the coefficients for each time series independently, which is also chosen as a benchmark method in \citet{BandedVAR}.
The setting of Case 1 is to validate that our proposed method is equivalent to the banded VAR method under the 1-D lattice structure.



Note that error standard deviation $\sigma_{e} = 0.01$ corresponds to very high signal and we would like to recover the true coefficient matrix exactly in this scenario.
On the other hand, error standard deviation $\sigma_{e} = 1$ corresponds to almost all noise, and hence almost all methods will perform relatively poorly.
Here we consider the $L_2$ error norm (spectral norm), i.e. $||\hat{A} - A||_2$, as the performance measure, where $\hat{A}$ is the estimated coefficient matrix, and $A$ is the true coefficient matrix.

\begin{figure}[H]
	\centering
	\includegraphics[scale=0.22]{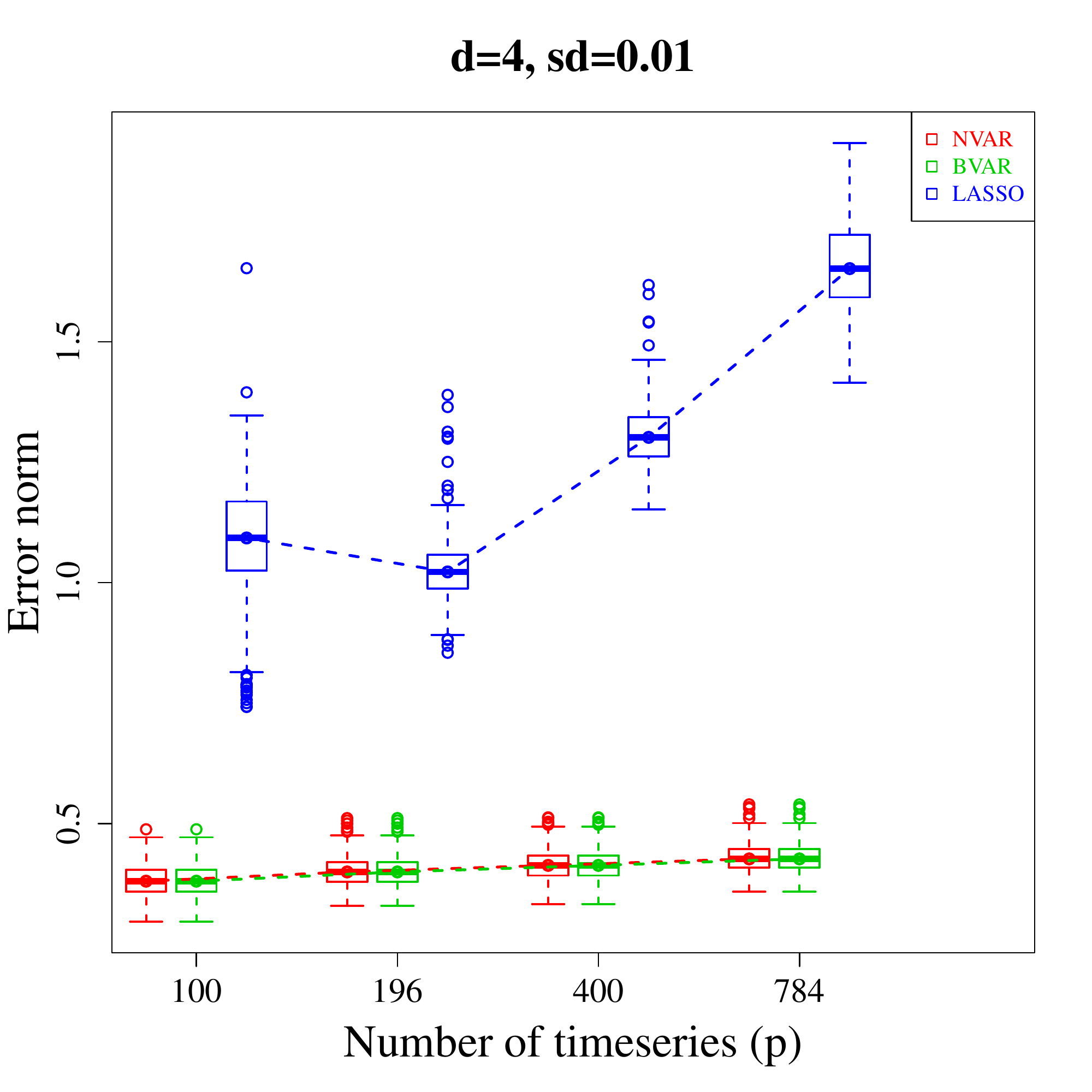}
	\includegraphics[scale=0.22]{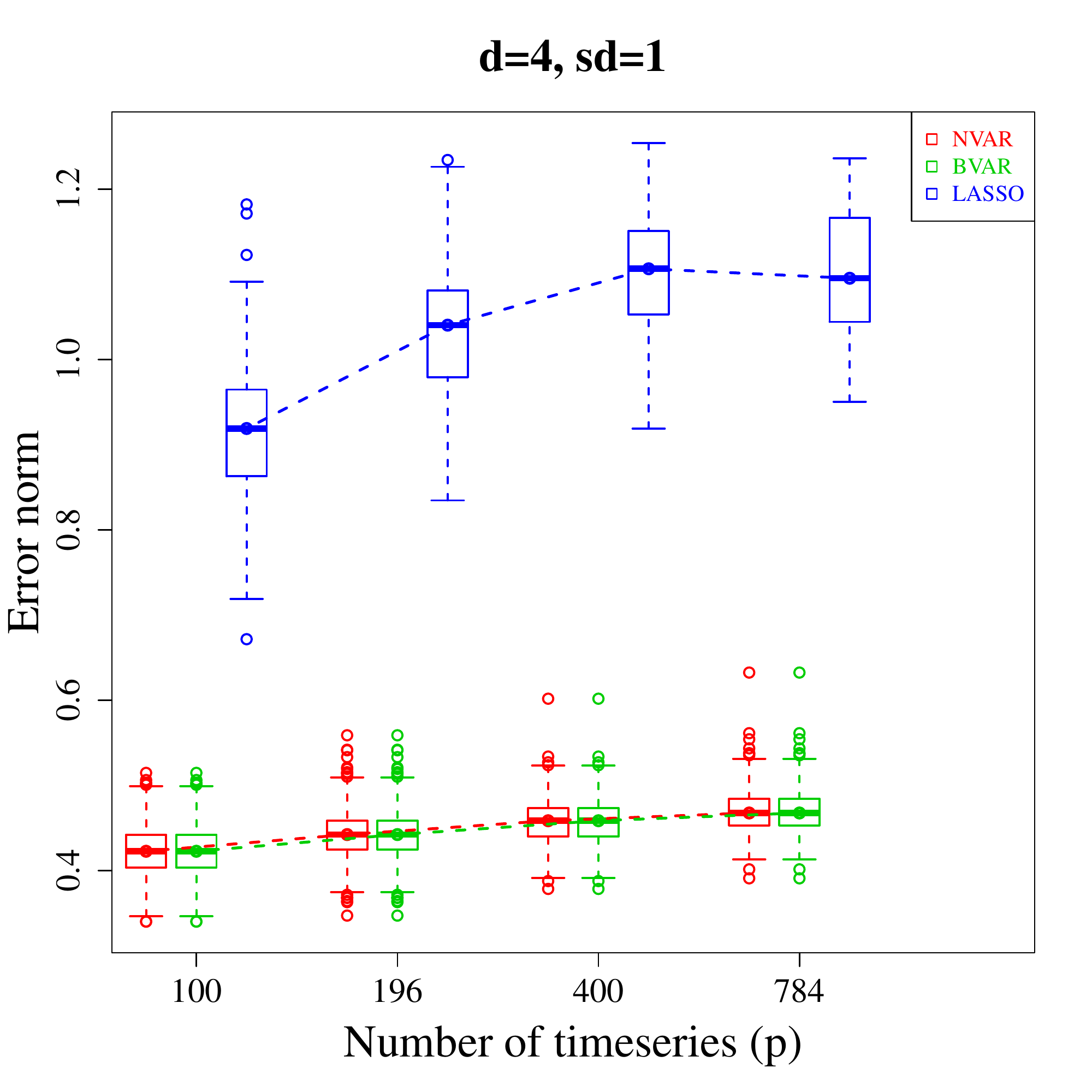}\\
	\includegraphics[scale=0.22]{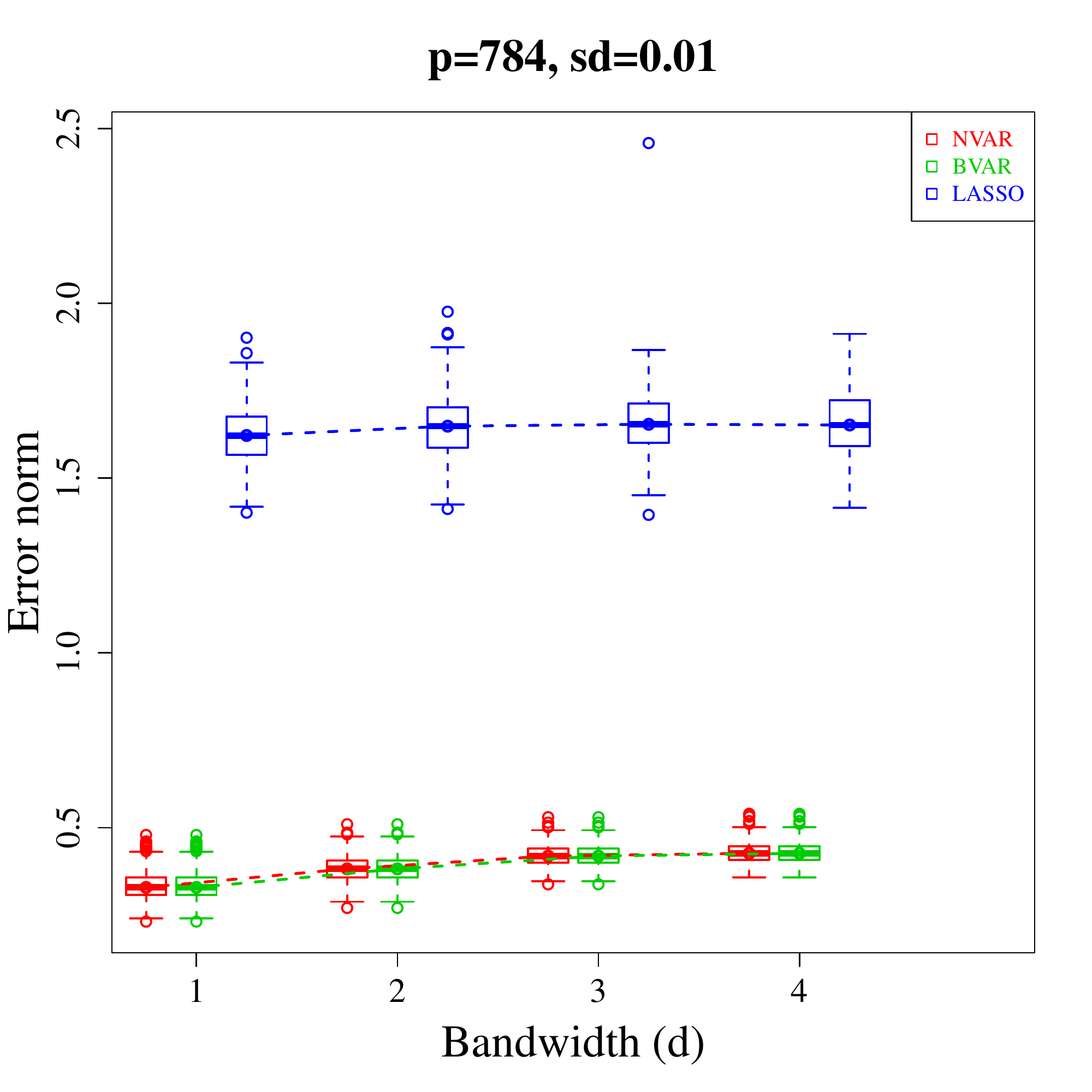}
	\includegraphics[scale=0.22]{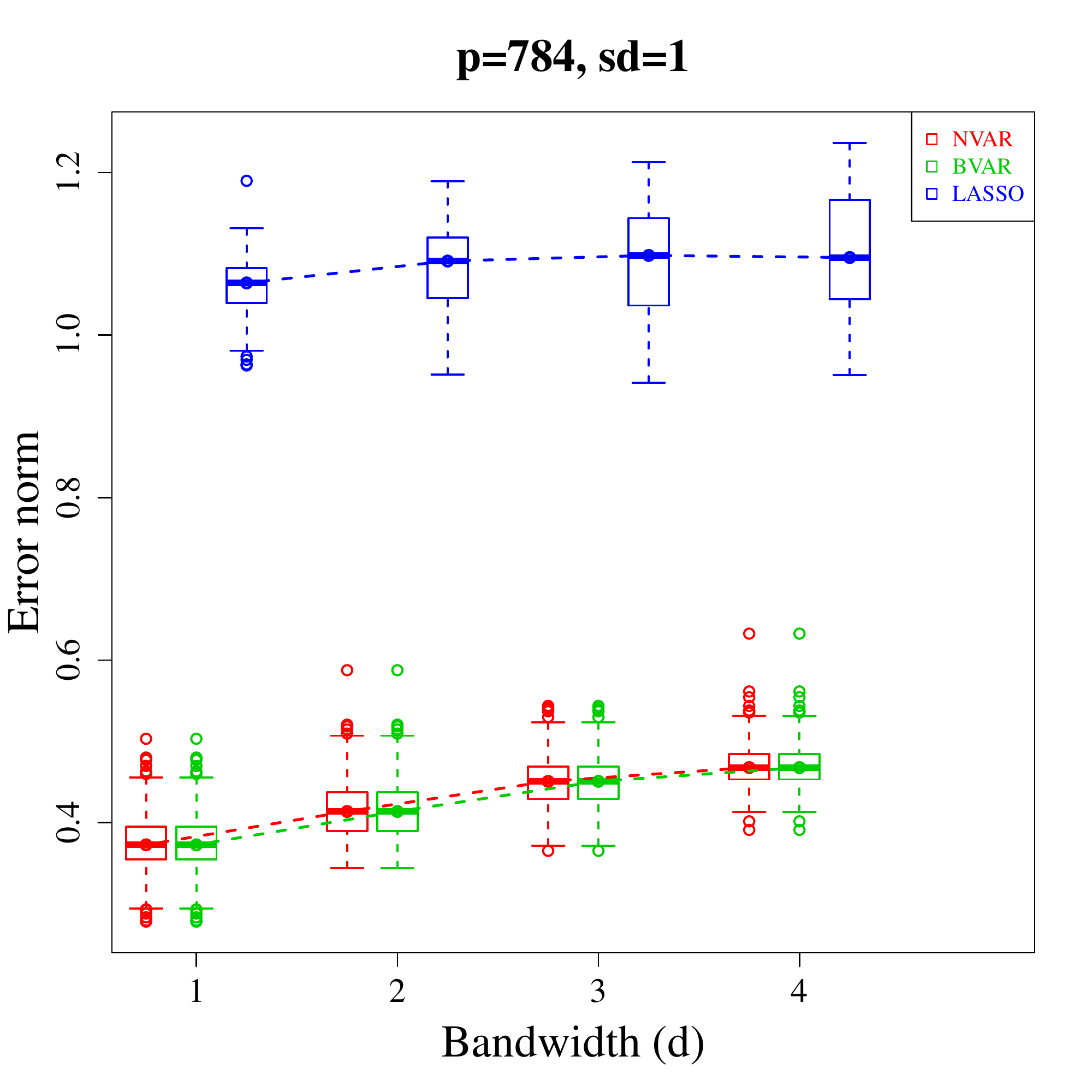}
	\caption{\label{fig: sim_perf_1_1}Performance comparison of methods with different number of time series ($p$), different bandwidth ($d$), and different error standard deviation (sd) in Case 1. The error norm is $||\hat{A}-A||_2$.}
\end{figure}

\begin{figure}[H]
	\centering
	\includegraphics[scale=0.22]{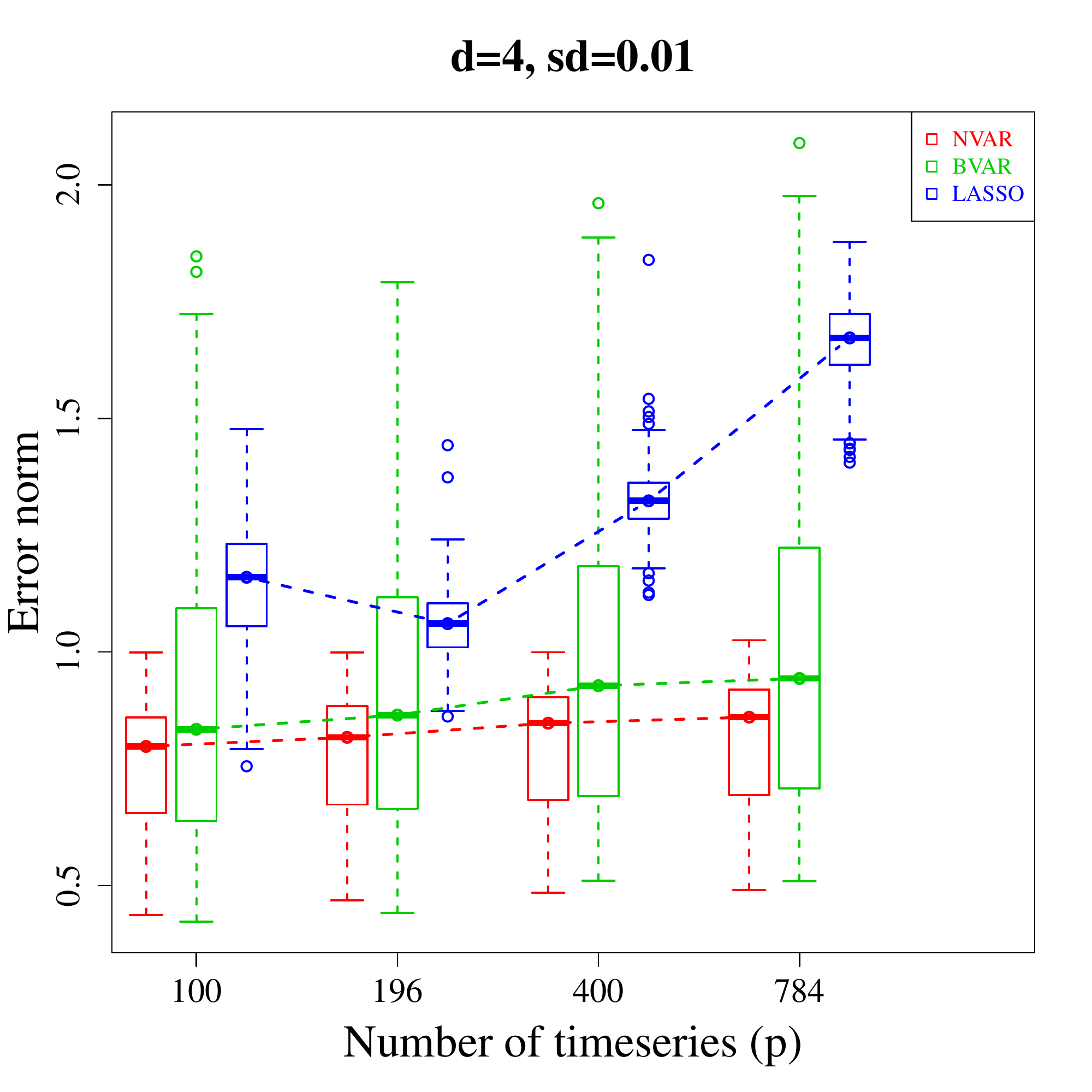}
	\includegraphics[scale=0.22]{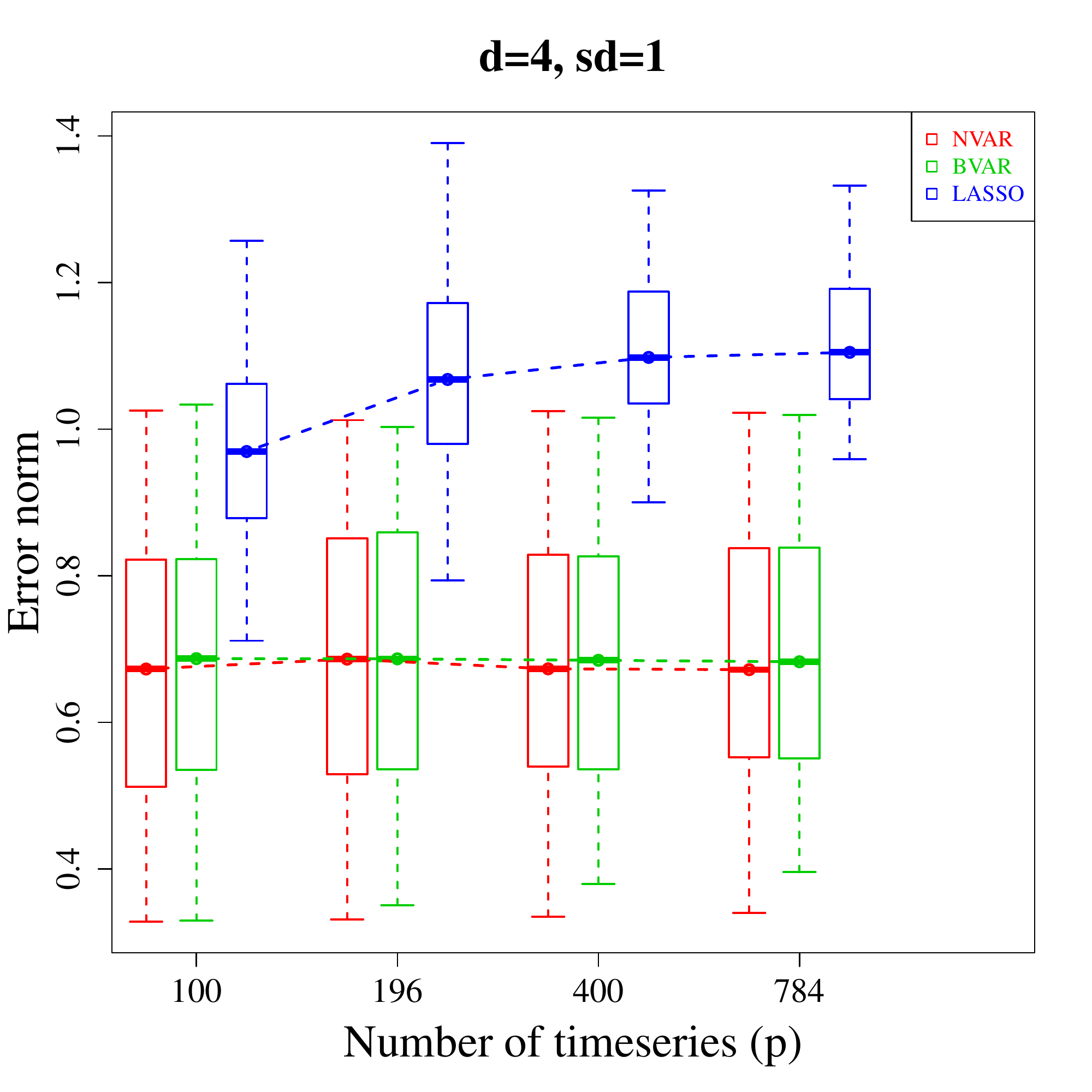}\\
	\includegraphics[scale=0.22]{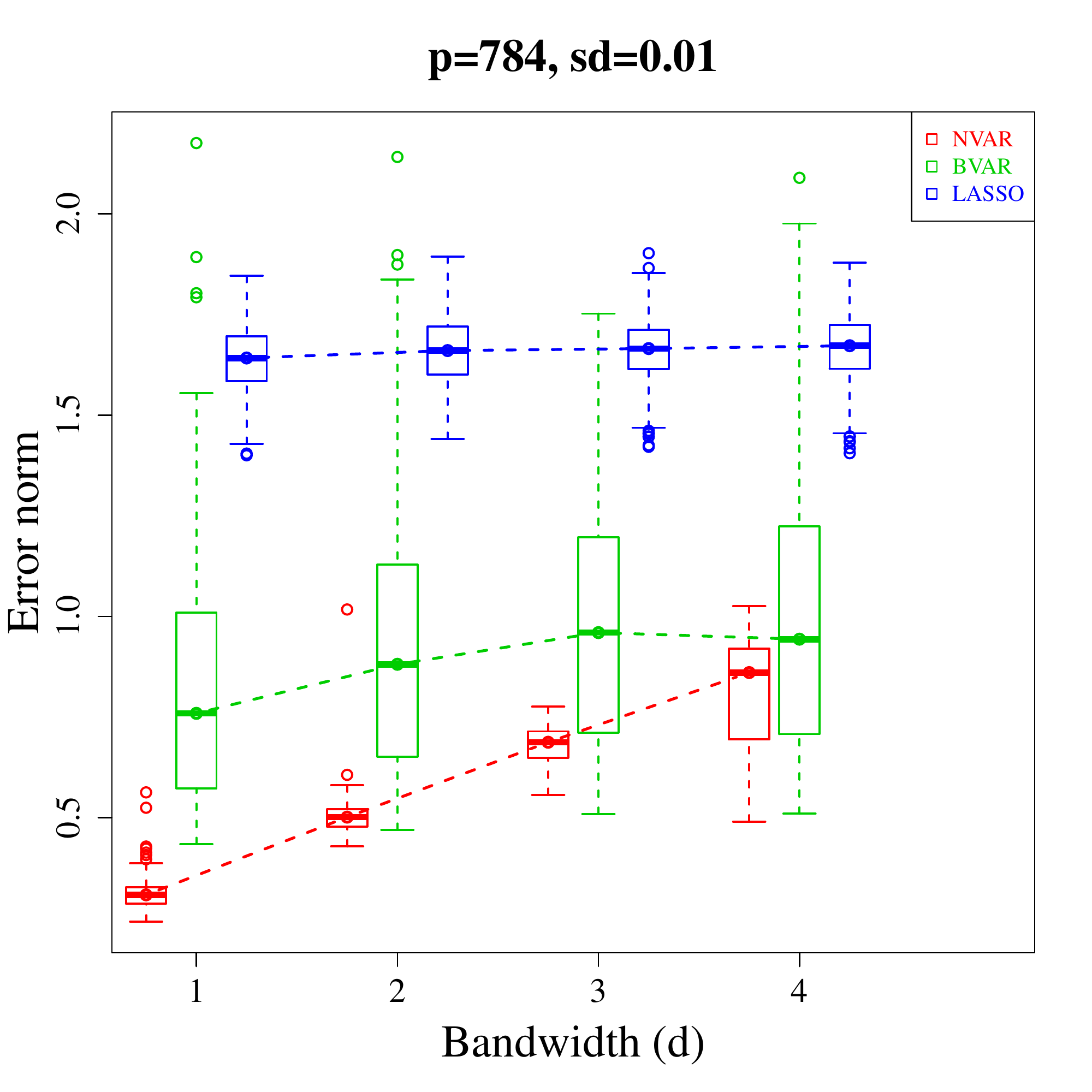}
	\includegraphics[scale=0.22]{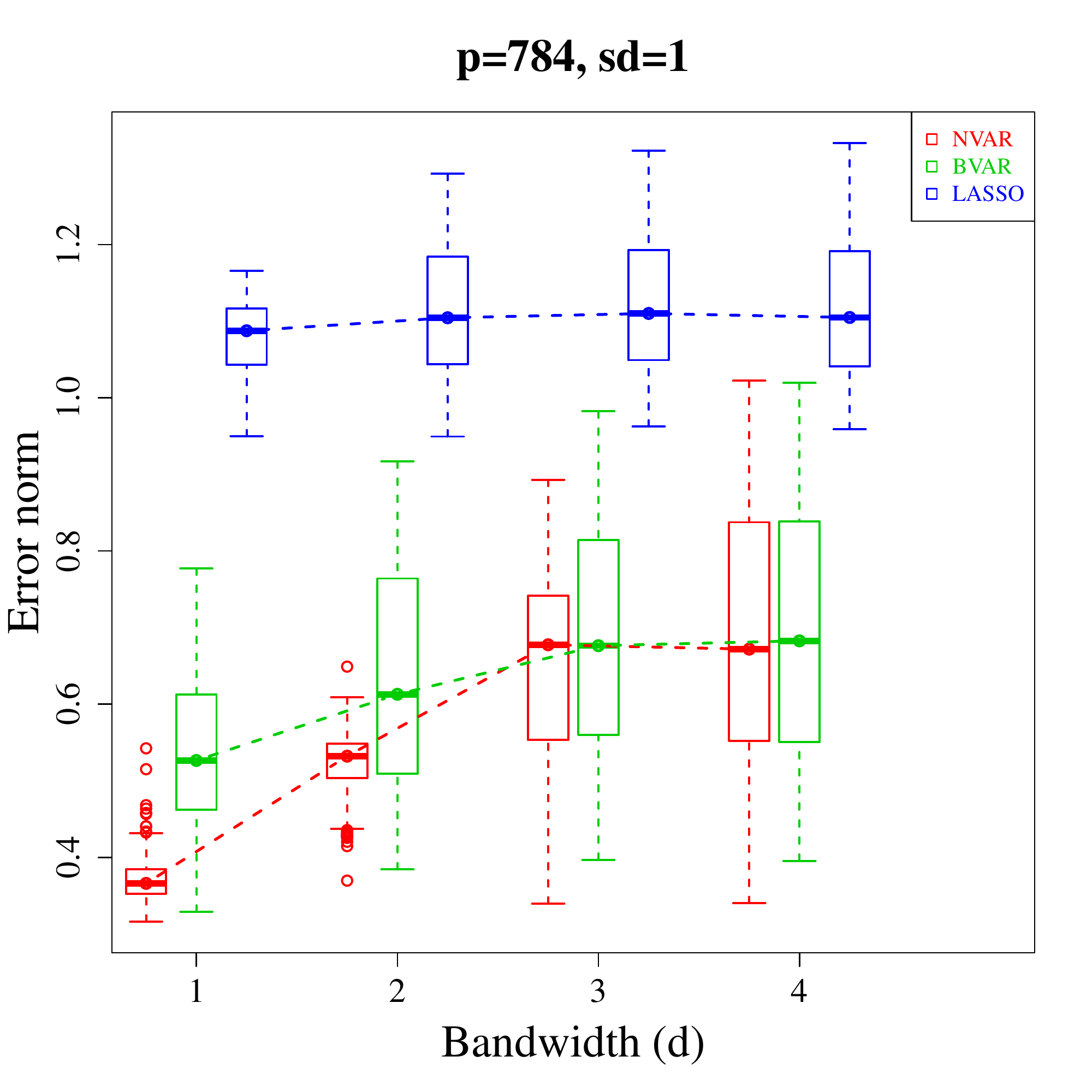}
	\vspace{-0.2cm}
	\caption{\label{fig: sim_perf_2_1}Performance comparison of methods with different number of time series ($p$), different bandwidth ($d$), and different error standard deviation (sd) in Case 2. The error norm is $||\hat{A}-A||_2$.
	}
\end{figure}

\begin{figure}[H]
	\centering
	\includegraphics[scale=0.25]{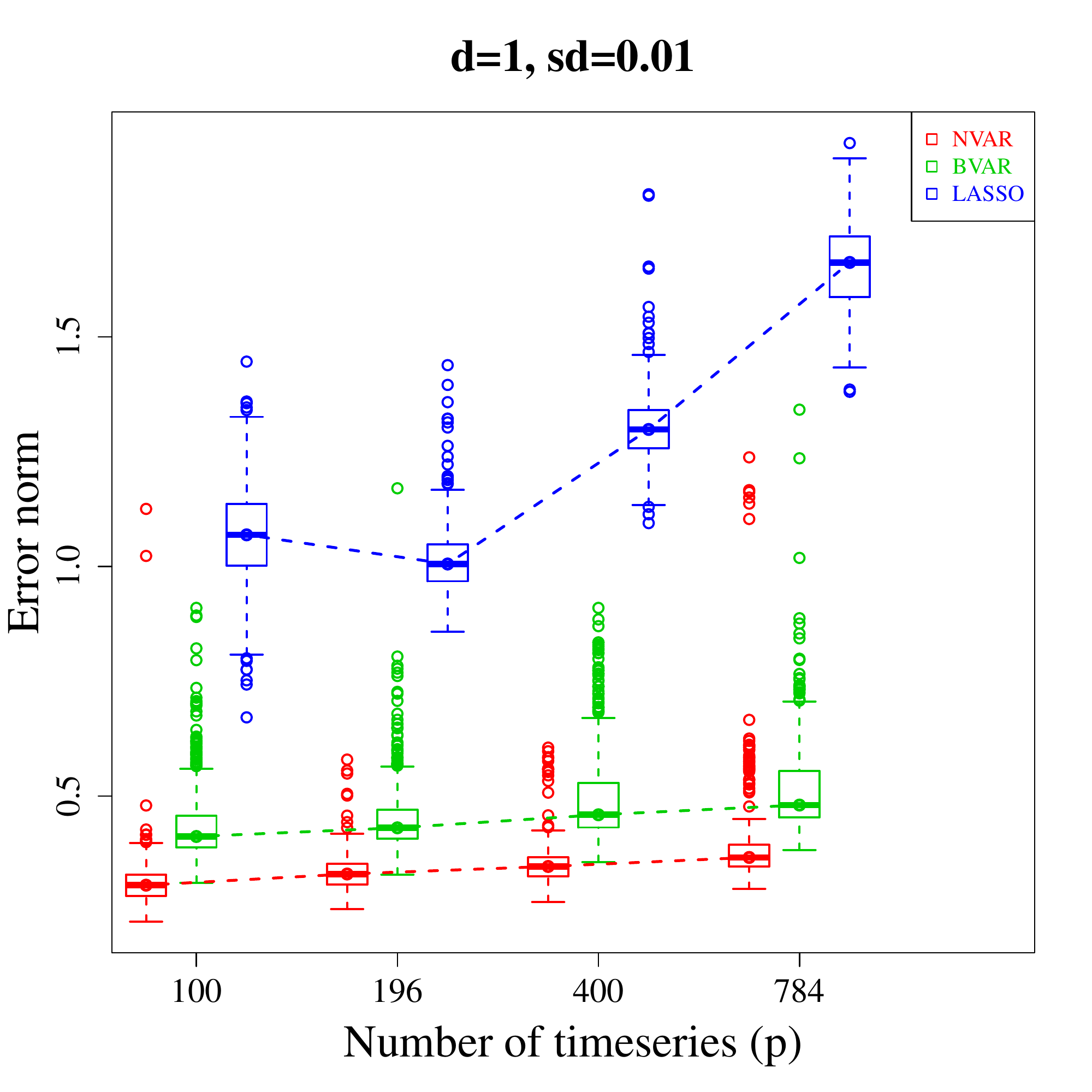}
	\includegraphics[scale=0.25]{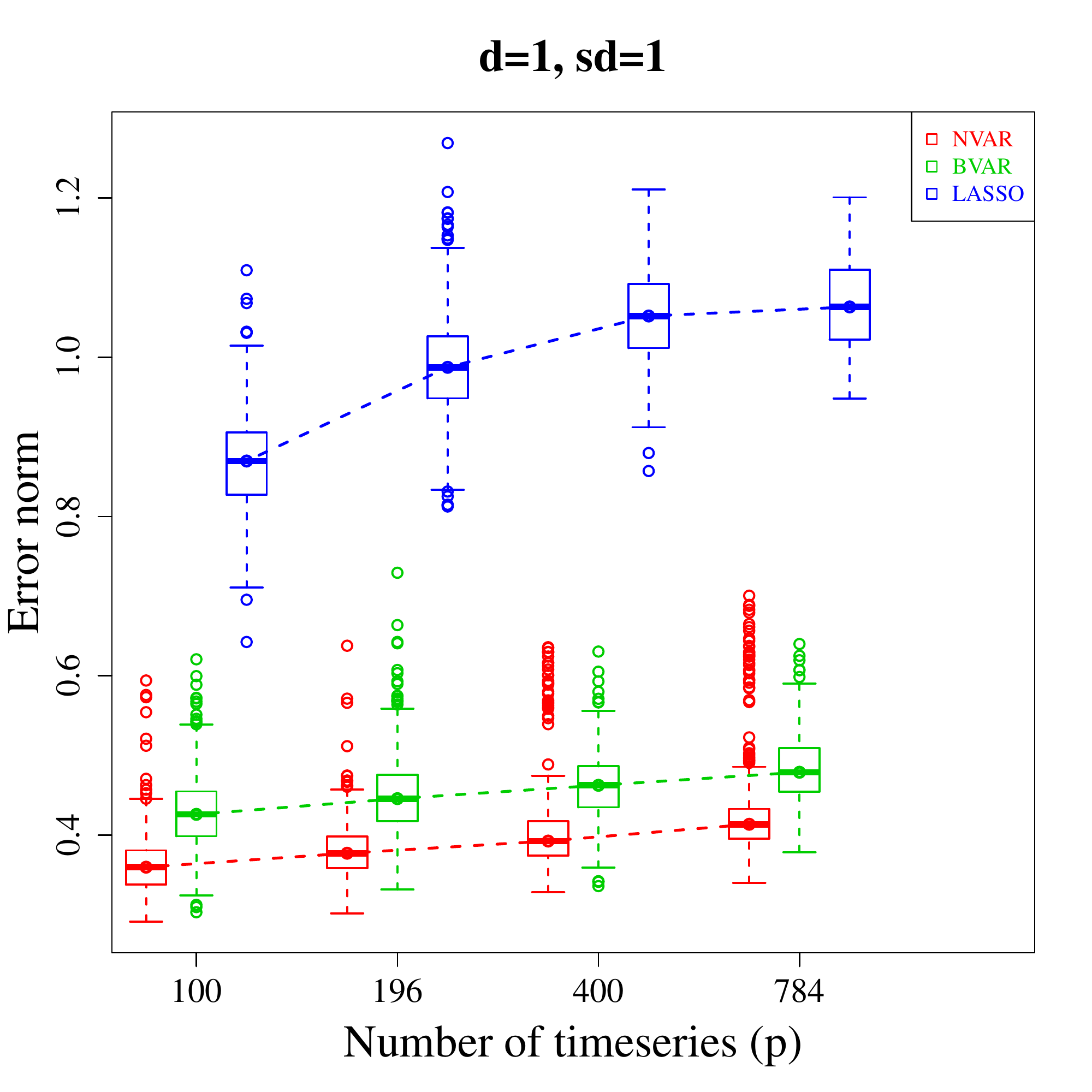}\\
	\includegraphics[scale=0.25]{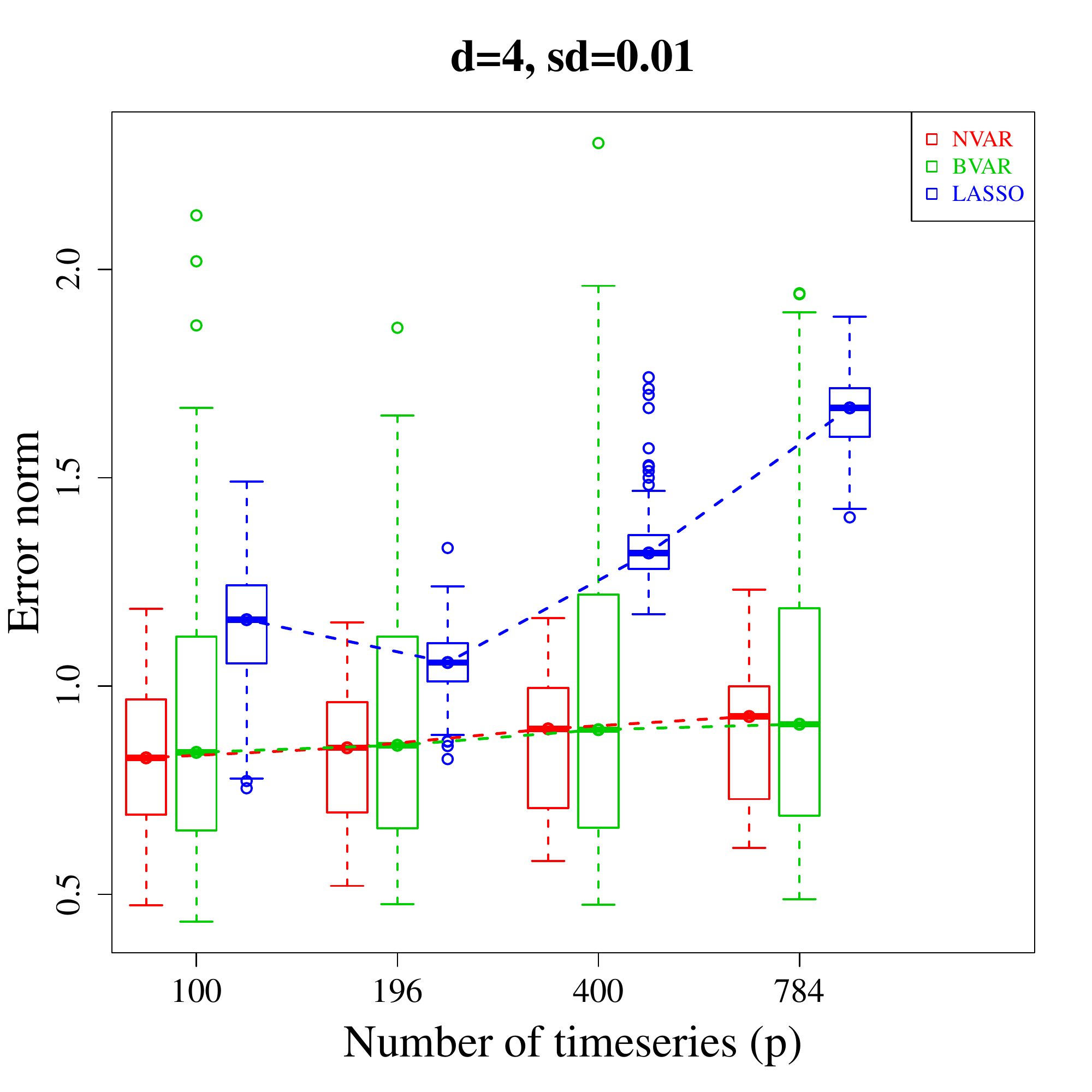}
	\includegraphics[scale=0.25]{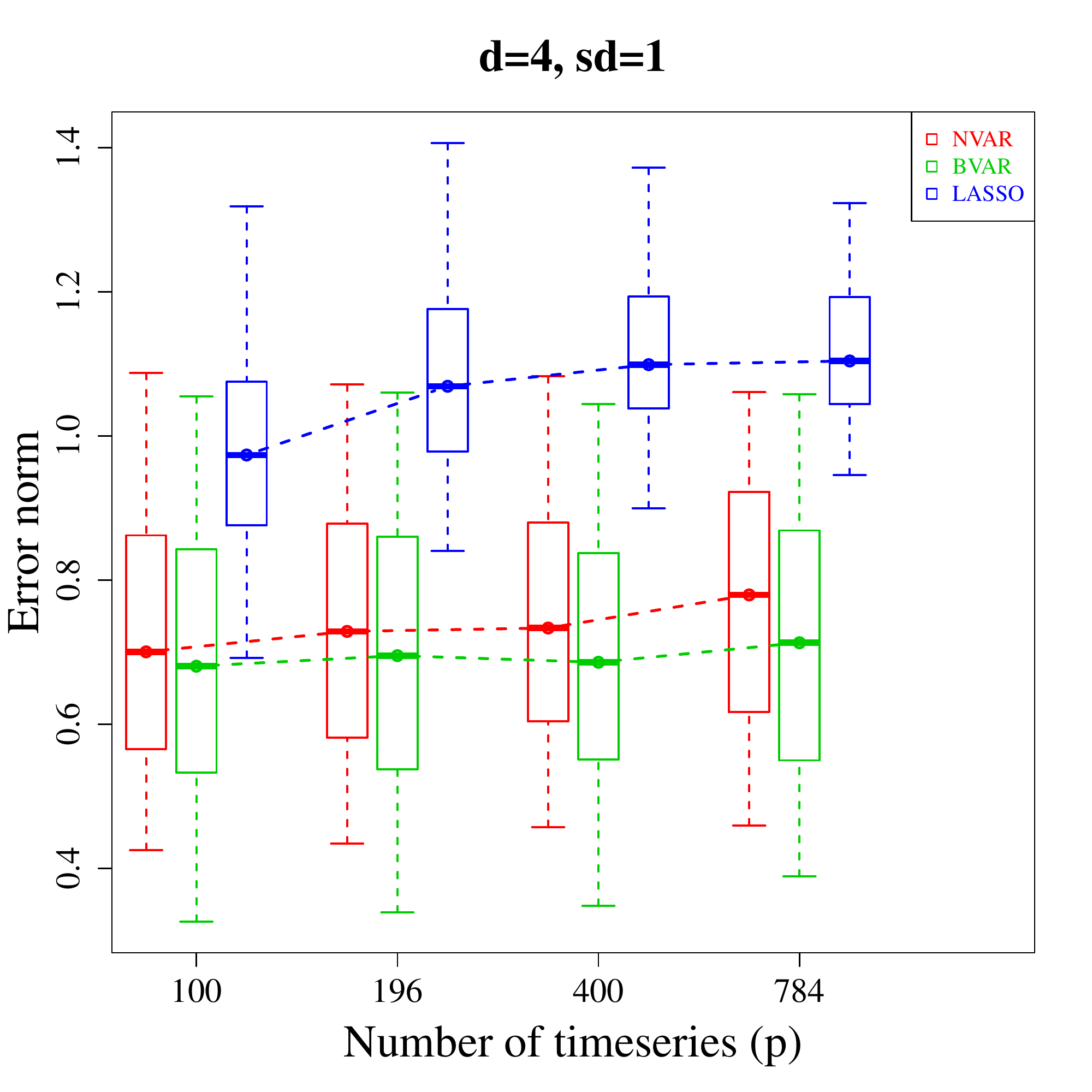}\\
	\includegraphics[scale=0.25]{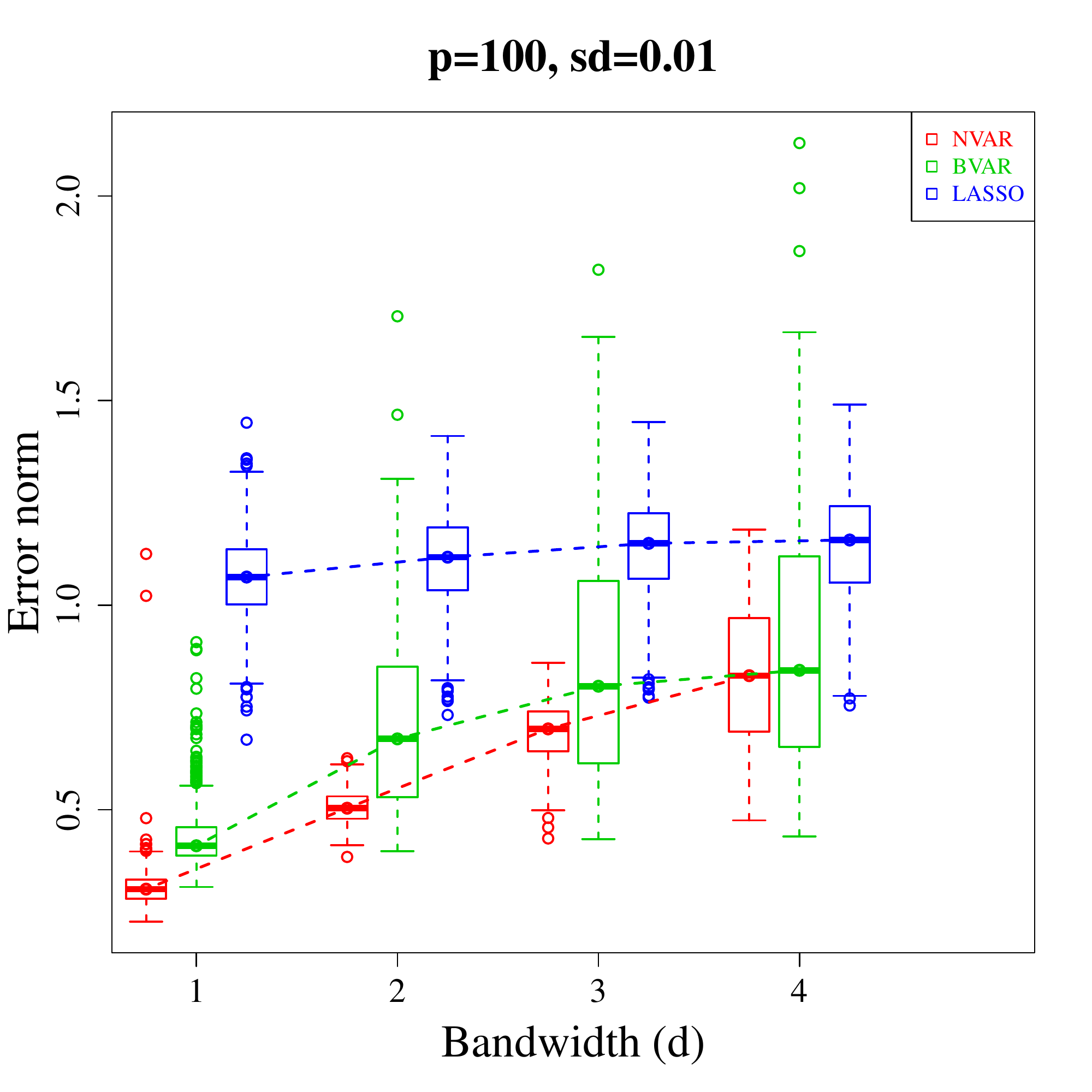}
	\includegraphics[scale=0.25]{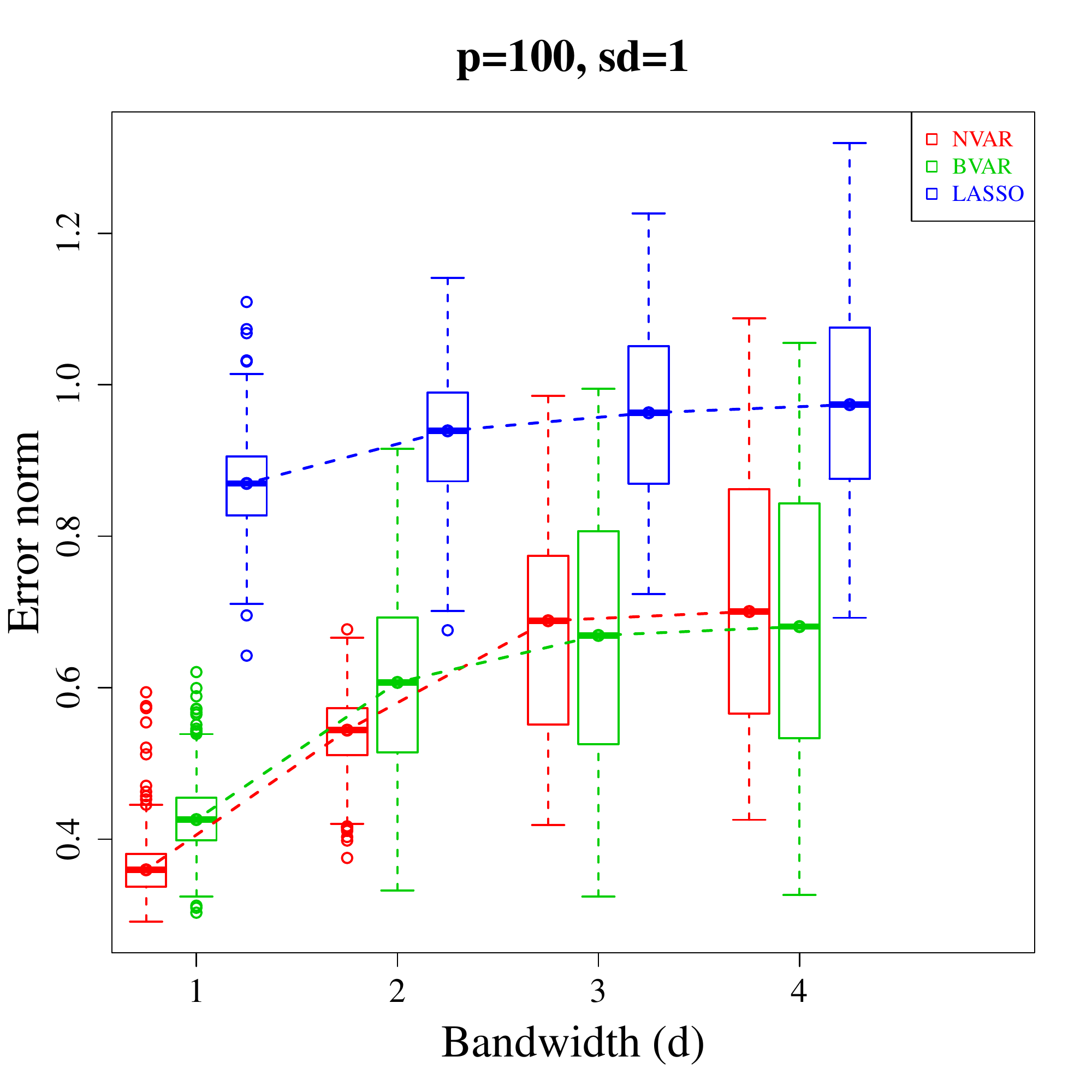}\\
	\includegraphics[scale=0.25]{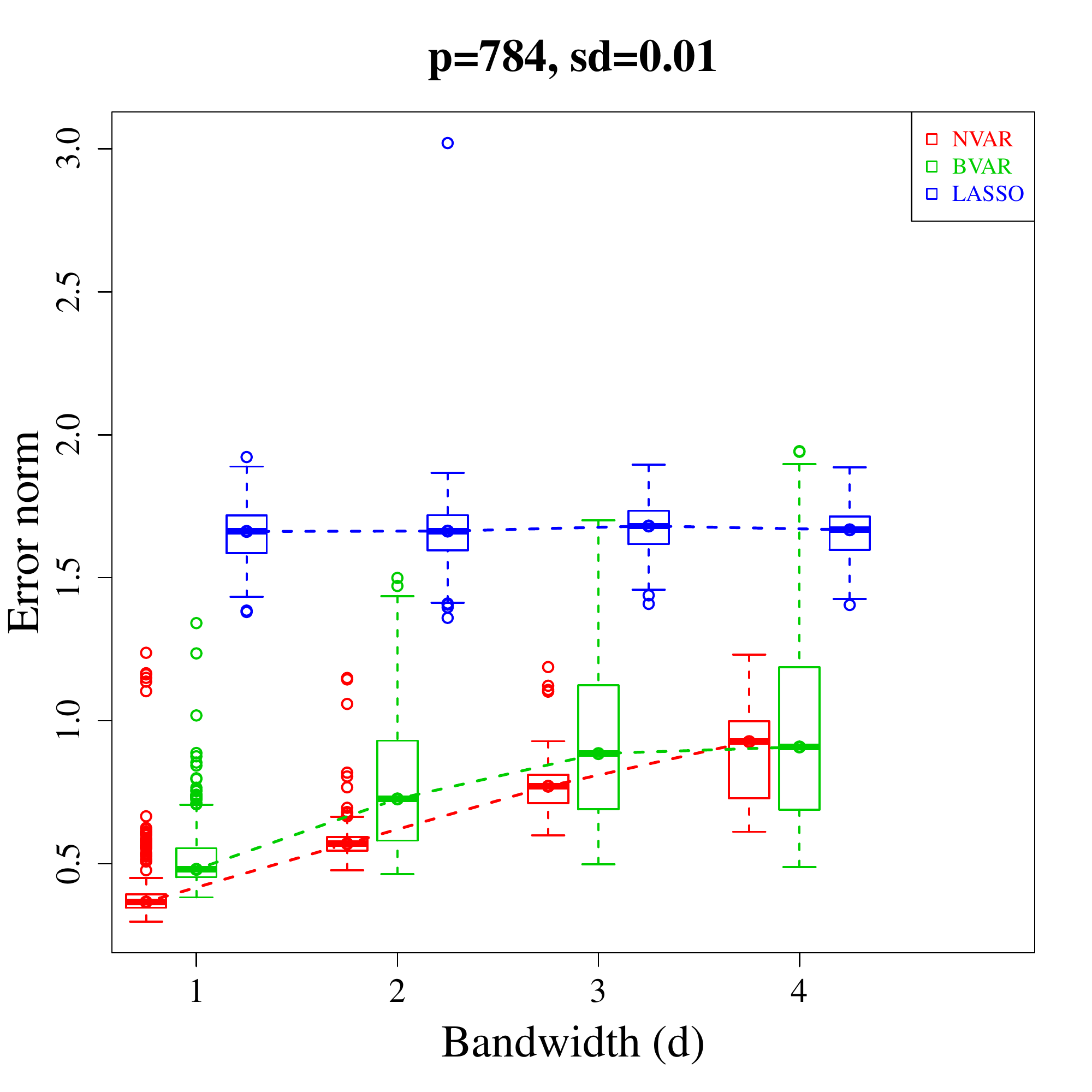}
	\includegraphics[scale=0.25]{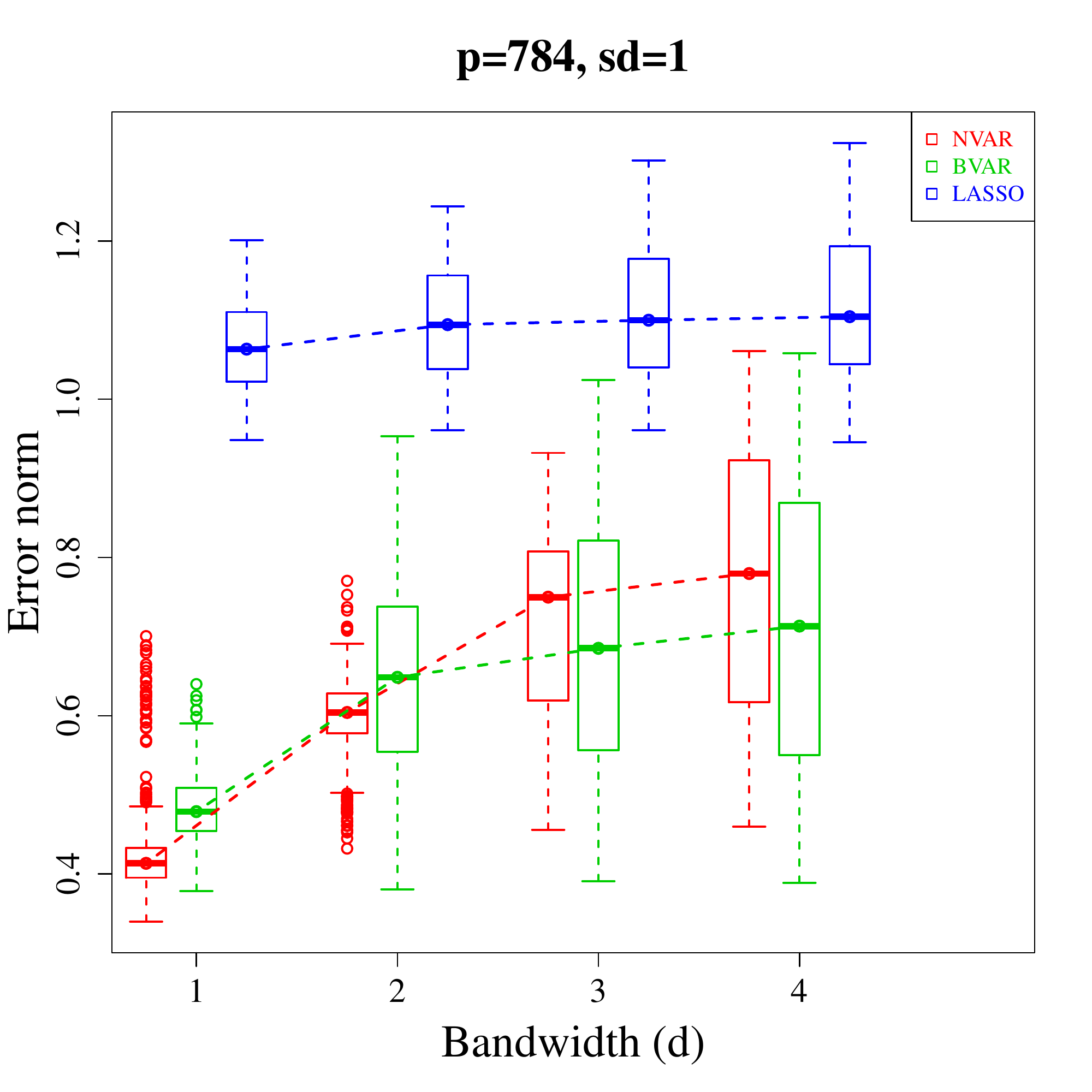}
	\caption{\label{fig: sim_perf_3_1}Performance comparison of methods with different number of time series ($p$), different bandwidth ($d$), and different error standard deviation (sd) in Case 3. The error norm is $||\hat{A}-A||_2$.
	}
\end{figure}

The simulation results are reported in Figures~\ref{fig: sim_perf_1_1}-\ref{fig: sim_perf_3_1} and Tables~\ref{tab: sim1}-\ref{tab: sim3}.
In Case 1, it is clear from Figures~\ref{fig: sim_perf_1_1} and Table~\ref{tab: sim1} that the banded VAR method and the proposed neighborhood VAR method coincide exactly,
and both perform much better than the LASSO method, irrespective of the noise variance.

In Case 2, the neighborhood VAR method outperforms the banded VAR method and the LASSO method, especially when the error variance is small and the bandwidth is small (Figures~\ref{fig: sim_perf_2_1}, Table~\ref{tab: sim2}).
We note here that the banded VAR method is adapted to the 2-D lattice case in a natural manner as the original paper \citet{BandedVAR} explicitly defines only a one-dimensional problem.
The neighborhood VAR method is significantly better than the banded VAR method at low error variance.
This is because banded VAR does not take the possibility of non-zero elements in the coefficient matrix far away from the main diagonal.
The banded VAR method and the neighborhood method provide a comparable performance when the error variance is high, and both outperform LASSO significantly.

In Case 3, the simulation results are reported in Figures~\ref{fig: sim_perf_3_1} and Table~\ref{tab: sim3}.
From these results, one can see that the neighborhood VAR method outperforms the banded VAR method in this case even for the normalized distances.

Recall from our simulation settings that we have re-scaled the spatial distances so that the average number of neighbors for any time series is roughly the same as in a 2-D lattice in order to facilitate a fair comparison with the banded VAR method.
This result indicates that for general spatial problems where the density of neighbors may be very different from that of a lattice process, the neighborhood VAR method can easily outperform the banded VAR method.
Note that when the number of time series grows large, or when the bandwidth grows large at fixed number of time series, the neighborhood VAR method and the banded VAR method become comparable.
It is clear that the neighborhood VAR method outperforms both the banded VAR method and the LASSO method significantly at low bandwidth,
and the neighborhood VAR method and the banded VAR method are comparable at high bandwidth.

We have also checked the performance of prediction accuracy for the proposed method in comparison with the BVAR and the LASSO methods.
Specifically, based on the fitted model, we conduct the one-step ahead prediction for 50 steps to calculate the mean squared prediction errors for the methods in comparison. The results show similar merits of the proposed NVAR method as shown in the estimated coefficient matrix $\hat{A}$ in comparison with the BVAR and LASSO methods, thus the results are omitted here.

\begin{table}[H]
	\caption{\label{tab: sim1}Case 1: Banded Structure. Means with their corresponding standard deviations in parentheses of the errors, and the frequency of estimated bandwidth in estimating coefficient matrix.}
	\centering
	\scriptsize
	\begin{tabular}{c|c|c|c|c|c|c|c|c|c|c|c|c|c|c}
		\hline \hline
		\multicolumn{15}{c}{\bf Error standard deviation = 1}\\
		\hline
		\multicolumn{2}{c|}{} & \multicolumn{6}{c|}{NVAR} & \multicolumn{6}{c|}{BVAR} & \multicolumn{1}{c}{LASSO} \\
		\hline
		\multicolumn{1}{m{0.2cm}}{\multirow{2}{*}{$p$}} & \multicolumn{1}{m{0.2cm}|}{\multirow{2}{*}{$d_0$}} &  \multicolumn{5}{c|}{\multirow{1}{*}{Est. bandwidth}} & \multicolumn{1}{m{1.2cm}|}{\multirow{2}{*}{\shortstack[l]{$L_2$ norm\\$||\hat{A}-A||_2$}}} & \multicolumn{5}{c|}{\multirow{1}{*}{Est. bandwidth}} & \multicolumn{1}{m{1.2cm}|}{\multirow{2}{*}{\shortstack[l]{$L_2$ norm\\$||\hat{A}-A||_2$}}}  & \multicolumn{1}{m{1.2cm}}{\multirow{2}{*}{\shortstack[l]{$L_2$ norm\\$||\hat{A}-A||_2$}}} \\
		\cline{3-7}\cline{9-13} 
		\multicolumn{2}{m{0.4cm}|}{\multirow{2}{*}{}} & 0 & 1 & 2 & 3 & 4 & & 0 & 1 & 2 & 3 & 4 &   &  \\
		\hline
		100 & 1 & 0 & 327 & 161 & 10 & 2 & 0.30(0.05) & 0 & 327 & 161 & 10 & 2 & 0.30(0.05) & 0.84(0.06) \\
		100 & 2 & 0 & 9 & 351 & 133 & 7 & 0.35(0.04) & 0 & 9 & 351 & 133 & 7 & 0.35(0.04) & 0.87(0.06) \\
		100 & 3 & 0 & 18 & 51 & 340 & 91 & 0.39(0.03) & 0 & 18 & 51 & 340 & 91 & 0.39(0.03) & 0.90(0.07) \\
		100 & 4 & 1 & 39 & 64 & 49 & 347 & 0.42(0.03) & 1 & 39 & 64 & 49 & 347 & 0.42(0.03) & 0.91(0.08) \\
		\hline
		196 & 1 & 0 & 251 & 232 & 16 & 1 & 0.32(0.04) & 0 & 251 & 232 & 16 & 1 & 0.32(0.04)& 0.96(0.06) \\
		196 & 2 & 0 & 5 & 295 & 183 & 17 & 0.37(0.04)& 0 & 5 & 295 & 183 & 17 & 0.37(0.04) & 0.99(0.06) \\
		196 & 3 & 0 & 7 & 33 & 312 & 148 & 0.41(0.03) & 0 & 7 & 33 & 312 & 148 & 0.41(0.03) & 1.02(0.07) \\
		196 & 4 & 0 & 19 & 48 & 51 & 382 & 0.44(0.03) & 0 & 19 & 48 & 51 & 382 & 0.44(0.03) & 1.03(0.07) \\
		\hline
		400 & 1 & 0 & 136 & 327 & 36 & 1 & 0.35(0.04) & 0 & 136 & 327 & 36 & 1 & 0.35(0.04) & 1.04(0.04) \\
		400 & 2 & 0 & 0 & 223 & 252 & 25 & 0.39(0.03) & 0 & 0 & 223 & 252 & 25 & 0.39(0.03) & 1.07(0.05) \\
		400 & 3 & 0 & 0 & 20 & 282 & 198 & 0.43(0.03) & 0 & 0 & 20 & 282 & 198 & 0.43(0.03) & 1.08(0.06) \\
		400 & 4 & 0 & 1 & 42 & 39 & 418 & 0.46(0.03) & 0 & 1 & 42 & 39 & 418 & 0.46(0.03) & 1.10(0.07) \\
		\hline
		784 & 1 & 0 & 52 & 395 & 49 & 4 & 0.38(0.03) & 0 & 52 & 395 & 49 & 4 & 0.38(0.03) & 1.06(0.03) \\
		784 & 2 & 0 & 0 & 156 & 297 & 47 & 0.42(0.04) & 0 & 0 & 156 & 297 & 47 & 0.42(0.04) & 1.08(0.05) \\
		784 & 3 & 0 & 2 & 14 & 242 & 242 & 0.45(0.03) & 0 & 2 & 14 & 242 & 242 & 0.45(0.03) & 1.09(0.06) \\
		784 & 4 & 0 & 1 & 33 & 50 & 416 & 0.47(0.03) & 0 & 1 & 33 & 50 & 416 & 0.47(0.03) & 1.10(0.07) \\
		\hline \hline
		\multicolumn{15}{c}{\bf Error standard deviation = 0.01}\\
        \hline
		\multicolumn{2}{c|}{} & \multicolumn{6}{c|}{NVAR} & \multicolumn{6}{c|}{BVAR} & \multicolumn{1}{c}{LASSO} \\
		\hline
		\multicolumn{1}{m{0.2cm}}{\multirow{2}{*}{$p$}} & \multicolumn{1}{m{0.2cm}|}{\multirow{2}{*}{$d_0$}} &  \multicolumn{5}{c|}{\multirow{1}{*}{Est. bandwidth}} & \multicolumn{1}{m{1.2cm}|}{\multirow{2}{*}{\shortstack[l]{$L_2$ norm\\$||\hat{A}-A||_2$}}} & \multicolumn{5}{c|}{\multirow{1}{*}{Est. bandwidth}} & \multicolumn{1}{m{1.2cm}|}{\multirow{2}{*}{\shortstack[l]{$L_2$ norm\\$||\hat{A}-A||_2$}}}  & \multicolumn{1}{m{1.2cm}}{\multirow{2}{*}{\shortstack[l]{$L_2$ norm\\$||\hat{A}-A||_2$}}} \\
		\cline{3-7}\cline{9-13}
		\multicolumn{2}{m{0.4cm}|}{\multirow{2}{*}{ }} & 0 & 1 & 2 & 3 & 4 & & 0 & 1 & 2 & 3 & 4 &   &  \\
		\hline
		100 & 1 & 0 & 278 & 207 & 13 & 2 & 0.24(0.16) & 0 & 278 & 207 & 13 & 2 & 0.24(0.06) & 1.01(0.11) \\
		100 & 2 & 0 & 0 & 314 & 168 & 18 & 0.30(0.05) & 0 & 0 & 314 & 168 & 18 & 0.30(0.05) & 1.06(0.10) \\
		100 & 3 & 0 & 0 & 0 & 330 & 170 & 0.35(0.04) & 0 & 0 & 0 & 330 & 170 & 0.35(0.04) & 1.08(0.12) \\
		100 & 4 & 0 & 0 & 1 & 24 & 475 & 0.38(0.03) & 0 & 0 & 1 & 24 & 475 & 0.38(0.03) & 1.09(0.12) \\
		\hline
		196 & 1 & 0 & 176 & 297 & 25 & 2 & 0.27(0.05) & 0 & 176 & 297 & 25 & 2 & 0.27(0.05) & 0.98(0.09) \\
		196 & 2 & 0 & 0 & 223 & 259 & 18 & 0.32(0.04) & 0 & 0 & 223 & 259 & 18 & 0.32(0.04) & 1.01(0.09) \\
		196 & 3 & 0 & 0 & 0 & 280 & 220 & 0.37(0.04) & 0 & 0 & 0 & 280 & 220 & 0.37(0.04) & 1.01(0.07) \\
		196 & 4 & 0 & 0 & 0 & 8 & 492 & 0.40(0.03) & 0 & 0 & 0 & 8 & 492 & 0.40(0.03) & 1.03(0.06) \\
		\hline
		400 & 1 & 0 & 68 & 398 & 31 & 3 & 0.31(0.04) & 0 & 68 & 398 & 31 & 3 & 0.31(0.04) & 1.28(0.09) \\
		400 & 2 & 0 & 0 & 103 & 356 & 41 & 0.36(0.04) & 0 & 0 & 103 & 356 & 41 & 0.36(0.04) & 1.28(0.07) \\
		400 & 3 & 0 & 0 & 0 & 174 & 326 & 0.40(0.03) & 0 & 0 & 0 & 174 & 326 & 0.40(0.03) & 1.30(0.07) \\
		400 & 4 & 0 & 0 & 0 & 1 & 499 & 0.41(0.03) & 0 & 0 & 0 & 1 & 499 & 0.41(0.03) & 1.31(0.07) \\
		\hline
		784 & 1 & 0 & 12 & 412 & 73 & 3 & 0.33(0.04) & 0 & 12 & 412 & 73 & 3 & 0.33(0.04) & 1.62(0.08) \\
		784 & 2 & 0 & 0 & 39 & 399 & 62 & 0.38(0.04) & 0 & 0 & 39 & 399 & 62 & 0.38(0.04) & 1.65(0.09) \\
		784 & 3 & 0 & 0 & 0 & 99 & 401 & 0.42(0.03) & 0 & 0 & 0 & 99 & 401 & 0.42(0.03) & 1.66(0.13) \\
		784 & 4 & 0 & 0 & 0 & 1 & 499 & 0.43(0.03) & 0 & 0 & 0 & 1 & 499 & 0.43(0.03) & 1.65(0.09) \\
		\hline \hline
	\end{tabular}
\end{table}

\begin{table}[H]
	\caption{\label{tab: sim2}Case 2: Block-banded Structure. Means with their corresponding standard deviations in parentheses of the errors, and the frequency of estimated bandwidth in estimating coefficient matrix.} 
	\centering
	\scriptsize
	\begin{tabular}{c|c|c|c|c|c|c|c|c|c|c|c|c|c|c}
		\hline \hline
		\multicolumn{15}{c}{\bf Error standard deviation = 1}\\
		\hline
		\multicolumn{2}{c|}{} & \multicolumn{6}{c|}{NVAR} & \multicolumn{6}{c|}{BVAR} & \multicolumn{1}{c}{LASSO} \\
		\hline
		\multicolumn{1}{m{0.2cm}}{\multirow{2}{*}{$p$}} & \multicolumn{1}{m{0.2cm}|}{\multirow{2}{*}{$d_0$}} &  \multicolumn{5}{c|}{\multirow{1}{*}{Est. bandwidth}} & \multicolumn{1}{m{1.2cm}|}{\multirow{2}{*}{\shortstack[l]{$L_2$ norm\\$||\hat{A}-A||_2$}}} & \multicolumn{5}{c|}{\multirow{1}{*}{Est. bandwidth}} & \multicolumn{1}{m{1.2cm}|}{\multirow{2}{*}{\shortstack[l]{$L_2$ norm\\$||\hat{A}-A||_2$}}}  & \multicolumn{1}{m{1.2cm}}{\multirow{2}{*}{\shortstack[l]{$L_2$ norm\\$||\hat{A}-A||_2$}}} \\
		\cline{3-7}\cline{9-13} 
		\multicolumn{2}{m{0.4cm}|}{\multirow{2}{*}{}} & 0 & 1 & 2 & 3 & 4 & & 0 & 1 & 2 & 3 & 4 &   &  \\
		\hline
		100 & 1 & 2 & 497 & 1 & 0 & 0 & 0.33(0.03) & 0 & 345 & 145 & 8 & 2 & 0.50(0.11) & 0.87(0.05) \\
		100 & 2 & 44 & 156 & 300 & 0 & 0 & 0.49(0.05) & 6 & 67 & 366 & 60 & 1 & 0.62(0.14) & 0.94(0.08) \\
		100 & 3 & 135 & 221 & 92 & 52 & 0 & 0.62(0.14) & 21 & 153 & 118 & 201 & 7 & 0.65(0.16) & 0.95(0.11) \\
		100 & 4 & 183 & 289 & 25 & 3 & 0 & 0.67(0.18) & 33 & 205 & 130 & 78 & 54 & 0.68(0.18) & 0.97(0.12) \\
		\hline
		196 & 1 & 0 & 495 & 5 & 0 & 0 & 0.34(0.03) & 0 & 273 & 208 & 18 & 1 & 0.51(0.11) & 0.99(0.07) \\
		196 & 2 & 21 & 169 & 310 & 0 & 0 & 0.50(0.04) & 0 & 47 & 366 & 82 & 5 & 0.63(0.14) & 1.05(0.09) \\
		196 & 3 & 92 & 240 & 109 & 59 & 0 & 0.63(0.14) & 4 & 121 & 125 & 237 & 13 & 0.67(0.16) & 1.07(0.11) \\
		196 & 4 & 152 & 308 & 38 & 2 & 0 & 0.68(0.19) & 12 & 186 & 155 & 91 & 56 & 0.69(0.18) & 1.08(0.12) \\
		\hline
		400 & 1 & 0 & 500 & 0 & 0 & 0 & 0.36(0.02)& 0 & 167 & 310 & 22 & 1 & 0.54(0.11) & 1.07(0.05) \\
		400 & 2 & 10 & 175 & 315 & 0 & 0 & 0.51(0.04) & 0 & 25 & 351 & 116 & 8 & 0.63(0.14) & 1.10(0.08) \\
		400 & 3 & 43 & 272 & 122 & 63 & 0 & 0.65(0.13) & 0 & 74 & 134 & 270 & 22 & 0.68(0.16) & 1.12(0.10) \\
		400 & 4 & 89 & 369 & 40 & 2 & 0 & 0.68(0.18) & 1 & 153 & 170 & 96 & 80 & 0.69(0.17) & 1.11(0.10) \\
		\hline
		784 & 1 & 0 & 498 & 2 & 0 & 0 & 0.37(0.03) & 0 & 84 & 381 & 34 & 1 & 0.54(0.10) & 1.08(0.05) \\
		784 & 2 & 1 & 180 & 319 & 0 & 0 & 0.52(0.04) & 0 & 12 & 297 & 174 & 17 & 0.64(0.14) & 1.11(0.08) \\
		784 & 3 & 21 & 284 & 120 & 75 & 0 & 0.65(0.12)& 0 & 37 & 140 & 270 & 53 & 0.69(0.15) & 1.13(0.13) \\
		784 & 4 & 53 & 386 & 57 & 4 & 0 & 0.69(0.17) & 0 & 89 & 179 & 126 & 106 & 0.70(0.17) & 1.12(0.09) \\
		\hline \hline
			\multicolumn{15}{c}{\bf Error standard deviation = 1}\\
		\hline
		\multicolumn{2}{c|}{} & \multicolumn{6}{c|}{NVAR} & \multicolumn{6}{c|}{BVAR} & \multicolumn{1}{c}{LASSO} \\
		\hline
		\multicolumn{1}{m{0.2cm}}{\multirow{2}{*}{$p$}} & \multicolumn{1}{m{0.2cm}|}{\multirow{2}{*}{$d_0$}} &  \multicolumn{5}{c|}{\multirow{1}{*}{Est. bandwidth}} & \multicolumn{1}{m{1.2cm}|}{\multirow{2}{*}{\shortstack[l]{$L_2$ norm\\$||\hat{A}-A||_2$}}} & \multicolumn{5}{c|}{\multirow{1}{*}{Est. bandwidth}} & \multicolumn{1}{m{1.2cm}|}{\multirow{2}{*}{\shortstack[l]{$L_2$ norm\\$||\hat{A}-A||_2$}}}  & \multicolumn{1}{m{1.2cm}}{\multirow{2}{*}{\shortstack[l]{$L_2$ norm\\$||\hat{A}-A||_2$}}} \\
		\cline{3-7}\cline{9-13} 
		\multicolumn{2}{m{0.4cm}|}{\multirow{2}{*}{}} & 0 & 1 & 2 & 3 & 4 & & 0 & 1 & 2 & 3 & 4 &   &  \\
		\hline
		100 & 1 & 0 & 498 & 2 & 0 & 0 & 0.26(0.04) & 0 & 0 & 12 & 70 & 418 & 0.75(0.23) & 1.06(0.10) \\
		100 & 2 & 0 & 20 & 480 & 0 & 0 & 0.45(0.04) & 0 & 0 & 11 & 66 & 423 & 0.83(0.27) & 1.11(0.11) \\
		100 & 3 & 0 & 14 & 211 & 275 & 0 & 0.62(0.05) & 0 & 0 & 3 & 54 & 443 & 0.86(0.28) & 1.13(0.13) \\
		100 & 4 & 0 & 23 & 223 & 186 & 68 & 0.76(0.12) & 0 & 0 & 2 & 25 & 473 & 0.88(0.28) & 1.14(0.13) \\
		\hline
		196 & 1 & 0 & 495 & 5 & 0 & 0 & 0.28(0.04) & 0 & 0 & 11 & 63 & 426 & 0.77(0.25)& 1.01(0.08) \\
		196 & 2 & 0 & 11 & 489 & 0 & 0 & 0.47(0.03) & 0 & 0 & 1 & 36 & 463 & 0.88(0.28) & 1.04(0.06)\\
		196 & 3 & 0 & 10 & 210 & 280 & 0 & 0.64(0.05) & 0 & 0 & 0 & 34 & 466 & 0.89(0.29) & 1.06(0.07) \\
		196 & 4 & 0 & 8 & 238 & 187 & 67 & 0.78(0.12) & 0 & 0 & 1 & 13 & 486 & 0.91(0.29) & 1.06(0.07) \\
		\hline
		400 & 1 & 0 & 498 & 2 & 0 & 0 & 0.29(0.04) & 0 & 0 & 0 & 42 & 458 & 0.82(0.28) & 1.28(0.08) \\
		400 & 2 & 0 & 4 & 495 & 0 & 1 & 0.49(0.04) & 0 & 0 & 0 & 13 & 487 & 0.89(0.30) & 1.31(0.06) \\
		400 & 3 & 0 & 4 & 202 & 294 & 0 & 0.67(0.05) & 0 & 0 & 0 & 11 & 489 & 0.93(0.30)& 1.32(0.06) \\
		400 & 4 & 0 & 5 & 217 & 199 & 79 & 0.80(0.12) & 0 & 0 & 0 & 2 & 498 & 0.96(0.32) & 1.33(0.06) \\
		\hline
		784 & 1 & 0 & 498 & 2 & 0 & 0 & 0.31(0.03) & 0 & 0 & 0 & 31 & 469 & 0.82(0.28) & 1.64(0.08) \\
		784 & 2 & 0 & 1 & 498 & 0 & 1 & 0.50(0.04) & 0 & 0 & 0 & 3 & 497 & 0.92(0.31) & 1.66(0.08) \\
		784 & 3 & 0 & 0 & 179 & 321 & 0 & 0.68(0.05) & 0 & 0 & 0 & 1 & 499 & 0.98(0.31) & 1.66(0.08) \\
		784 & 4 & 0 & 2 & 195 & 223 & 80 & 0.82(0.12) & 0 & 0 & 0 & 0 & 500 & 0.99(0.33) & 1.71(0.092) \\
		\hline \hline
	\end{tabular}
\end{table}

\begin{table}[H]
	\caption{\label{tab: sim3}Case 3: 2-D Spatial Structure. Means with their corresponding standard deviations in parentheses of the errors, and the frequency of estimated bandwidth in estimating coefficient matrix.} 
	\centering
	\scriptsize
	\begin{tabular}{c|c|c|c|c|c|c|c|c|c|c|c|c|c|c}
		\hline \hline
		\multicolumn{15}{c}{\bf Error standard deviation = 1}\\
		\hline
		\multicolumn{2}{c|}{} & \multicolumn{6}{c|}{NVAR} & \multicolumn{6}{c|}{BVAR} & \multicolumn{1}{c}{LASSO} \\
		\hline
		\multicolumn{1}{m{0.2cm}}{\multirow{2}{*}{$p$}} & \multicolumn{1}{m{0.2cm}|}{\multirow{2}{*}{$d_0$}} &  \multicolumn{5}{c|}{\multirow{1}{*}{Est. bandwidth}} & \multicolumn{1}{m{1.2cm}|}{\multirow{2}{*}{\shortstack[l]{$L_2$ norm\\$||\hat{A}-A||_2$}}} & \multicolumn{5}{c|}{\multirow{1}{*}{Est. bandwidth}} & \multicolumn{1}{m{1.2cm}|}{\multirow{2}{*}{\shortstack[l]{$L_2$ norm\\$||\hat{A}-A||_2$}}}  & \multicolumn{1}{m{1.2cm}}{\multirow{2}{*}{\shortstack[l]{$L_2$ norm\\$||\hat{A}-A||_2$}}} \\
		\cline{3-7}\cline{9-13} 
		\multicolumn{2}{m{0.4cm}|}{\multirow{2}{*}{}} & 0 & 1 & 2 & 3 & 4 & & 0 & 1 & 2 & 3 & 4 &   &  \\
		\hline
		100 & 1 & 0 & 494 & 6 & 0 & 0 & 0.36(0.04) & 0 & 33 & 93 & 107 & 267 & 0.43(0.05) & 0.87(0.06) \\
		100 & 2 & 0 & 139 & 361 & 0 & 0 & 0.54(0.05) & 6 & 85 & 89 & 81 & 239 & 0.60(0.12) & 0.93(0.08) \\
		100 & 3 & 0 & 283 & 174 & 43 & 0 & 0.67(0.13) & 24 & 170 & 93 & 107 & 106 & 0.66(0.17) & 0.96(0.11) \\
		100 & 4 & 0 & 394 & 101 & 5 & 0 & 0.71(0.17) & 47 & 240 & 127 & 61 & 25 & 0.68(0.18) & 0.98(0.13) \\
		\hline
		196 & 1 & 0 & 497 & 3 & 0 & 0 & 0.38(0.04) & 0 & 18 & 87 & 90 & 305 & 0.45(0.05) & 0.99(0.07) \\
		196 & 2 & 0 & 119 & 381 & 0 & 0 & 0.55(0.05) & 2 & 61 & 104 & 72 & 261 & 0.60(0.12) & 1.03(0.09) \\
		196 & 3 & 0 & 251 & 210 & 39 & 0 & 0.68(0.14) & 2 & 141 & 136 & 95 & 126 & 0.67(0.17) & 1.05(0.11) \\
		196 & 4 & 0 & 329 & 168 & 3 & 0 & 0.73(0.17) & 10 & 215 & 157 & 87 & 31 & 0.69(0.18) & 1.08(0.12) \\
		\hline
		400 & 1 & 0 & 473 & 27 & 0 & 0 & 0.40(0.05) & 0 & 6 & 76 & 115 & 303 & 0.46(0.04) & 1.05(0.06) \\
		400 & 2 & 0 & 94 & 406 & 0 & 0 & 0.57(0.05) & 0 & 26 & 93 & 84 & 297 & 0.62(0.12) & 1.09(0.08) \\
		400 & 3 & 0 & 189 & 245 & 66 & 0 & 0.71(0.13) & 0 & 93 & 135 & 91 & 181 & 0.69(0.17) & 1.11(0.10) \\
		400 & 4 & 0 & 281 & 215 & 4 & 0 & 0.74(0.16) & 2 & 170 & 166 & 103 & 59 & 0.70(0.17) & 1.11(0.10) \\
		\hline
		784 & 1 & 0 & 469 & 31 & 0 & 0 & 0.43(0.06) & 0 & 0 & 63 & 107 & 330 & 0.48(0.05) & 1.06(0.05) \\
		784 & 2 & 0 & 68 & 432 & 0 & 0 & 0.60(0.05) & 0 & 15 & 67 & 79 & 339 & 0.64(0.12) & 1.10(0.07) \\
		784 & 3 & 0 & 155 & 269 & 76 & 0 & 0.72(0.12) & 0 & 58 & 132 & 101 & 209 & 0.69(0.16) & 1.15(0.98) \\
		784 & 4 & 0 & 217 & 278 & 5 & 0 & 0.77(0.17) & 0 & 102 & 181 & 111 & 106 & 0.71(0.18) & 1.12(0.09) \\
		\hline \hline
			\multicolumn{15}{c}{\bf Error standard deviation = 1}\\
		\hline
		\multicolumn{2}{c|}{} & \multicolumn{6}{c|}{NVAR} & \multicolumn{6}{c|}{BVAR} & \multicolumn{1}{c}{LASSO} \\
		\hline
		\multicolumn{1}{m{0.2cm}}{\multirow{2}{*}{$p$}} & \multicolumn{1}{m{0.2cm}|}{\multirow{2}{*}{$d_0$}} &  \multicolumn{5}{c|}{\multirow{1}{*}{Est. bandwidth}} & \multicolumn{1}{m{1.2cm}|}{\multirow{2}{*}{\shortstack[l]{$L_2$ norm\\$||\hat{A}-A||_2$}}} & \multicolumn{5}{c|}{\multirow{1}{*}{Est. bandwidth}} & \multicolumn{1}{m{1.2cm}|}{\multirow{2}{*}{\shortstack[l]{$L_2$ norm\\$||\hat{A}-A||_2$}}}  & \multicolumn{1}{m{1.2cm}}{\multirow{2}{*}{\shortstack[l]{$L_2$ norm\\$||\hat{A}-A||_2$}}} \\
		\cline{3-7}\cline{9-13} 
		\multicolumn{2}{m{0.4cm}|}{\multirow{2}{*}{}} & 0 & 1 & 2 & 3 & 4 & & 0 & 1 & 2 & 3 & 4 &   &  \\
		\hline
		100 & 1 & 0 & 497 & 1 & 0 & 2 & 0.31(0.06) & 0 & 0 & 7 & 70 & 423 & 0.44(0.08) & 1.07(0.11) \\
		100 & 2 & 0 & 16 & 484 & 0 & 0 & 0.51(0.04) & 0 & 0 & 1 & 33 & 466 & 0.71(0.21) & 1.11(0.12) \\
		100 & 3 & 0 & 7 & 264 & 229 & 0 & 0.69(0.07) & 0 & 0 & 1 & 23 & 476 & 0.84(0.26) & 1.14(0.13) \\
		100 & 4 & 0 & 11 & 283 & 172 & 34 & 0.83(0.15) & 0 & 0 & 3 & 40 & 457 & 0.89(0.30) & 1.14(0.13) \\
		\hline
		196 & 1 & 0 & 495 & 5 & 0 & 0 & 0.33(0.04) & 0 & 0 & 3 & 84 & 413 & 0.45(0.08) & 1.01(0.07) \\
		196 & 2 & 0 & 5 & 493 & 1 & 1 & 0.53(0.05) & 0 & 0 & 1 & 21 & 478 & 0.73(0.22) & 1.04(0.07) \\
		196 & 3 & 0 & 2 & 239 & 258 & 1 & 0.72(0.07) & 0 & 0 & 0 & 14 & 486 & 0.87(0.26) & 1.06(0.07) \\
		196 & 4 & 0 & 1 & 267 & 196 & 36 & 0.84(0.15) & 0 & 0 & 0 & 17 & 483 & 0.91(0.29) & 1.06(0.07) \\
		\hline
		400 & 1 & 0 & 488 & 12 & 0 & 0 & 0.35(0.05) & 0 & 0 & 0 & 44 & 456 & 0.50(0.10) & 1.30(0.08) \\
		400 & 2 & 0 & 2 & 496 & 1 & 1 & 0.55(0.05) & 0 & 0 & 0 & 9 & 491 & 0.77(0.23) & 1.32(0.07) \\
		400 & 3 & 0 & 0 & 224 & 274 & 2 & 0.73(0.07) & 0 & 0 & 0 & 5 & 495 & 0.89(0.27) & 1.32(0.07) \\
		400 & 4 & 0 & 0 & 230 & 218 & 52 & 0.86(0.15) & 0 & 0 & 0 & 7 & 493 & 0.96(0.34) & 1.32(0.07) \\
		\hline
		784 & 1 & 0 & 461 & 33 & 0 & 6 & 0.39(0.10) & 0 & 0 & 0 & 25 & 475 & 0.52(0.10) & 1.65(0.10) \\
		784 & 2 & 0 & 0 & 494 & 3 & 3 & 0.58(0.06) & 0 & 0 & 0 & 3 & 497 & 0.77(0.22) & 1.65(0.11) \\
		784 & 3 & 0 & 0 & 199 & 296 & 5 & 0.77(0.08) & 0 & 0 & 0 & 1 & 499 & 0.92(0.28) & 1.67(0.09) \\
		784 & 4 & 0 & 0 & 217 & 236 & 47 & 0.88(0.15) & 0 & 0 & 0 & 1 & 499 & 0.97(0.33) & 1.66(0.09) \\
		\hline \hline
	\end{tabular}
\end{table}

\section{Case Study of Stream Nitrogen Data}
Excess nitrogen is one of the most important causes of impairment for rivers and streams (see \citet{EPA2000}).
Previous studies also showed that nitrogen and phosphorus loads are important driving factors of Harmful algal blooms (HABs) (see \citet{Paerl2001}).
For urban and urbanizing watersheds, less developed agricultural and low-density residential (suburban/exurban) areas contribute most in terms of annual loads of nitrogen, mainly through sewage and fertilizers (see \citet{Shields2008}). Thus, it is of great importance to understand the complication and prediction of nitrogen of multiple streams.

\begin{figure}[H]
	\centering
	\includegraphics[width=0.7\linewidth]{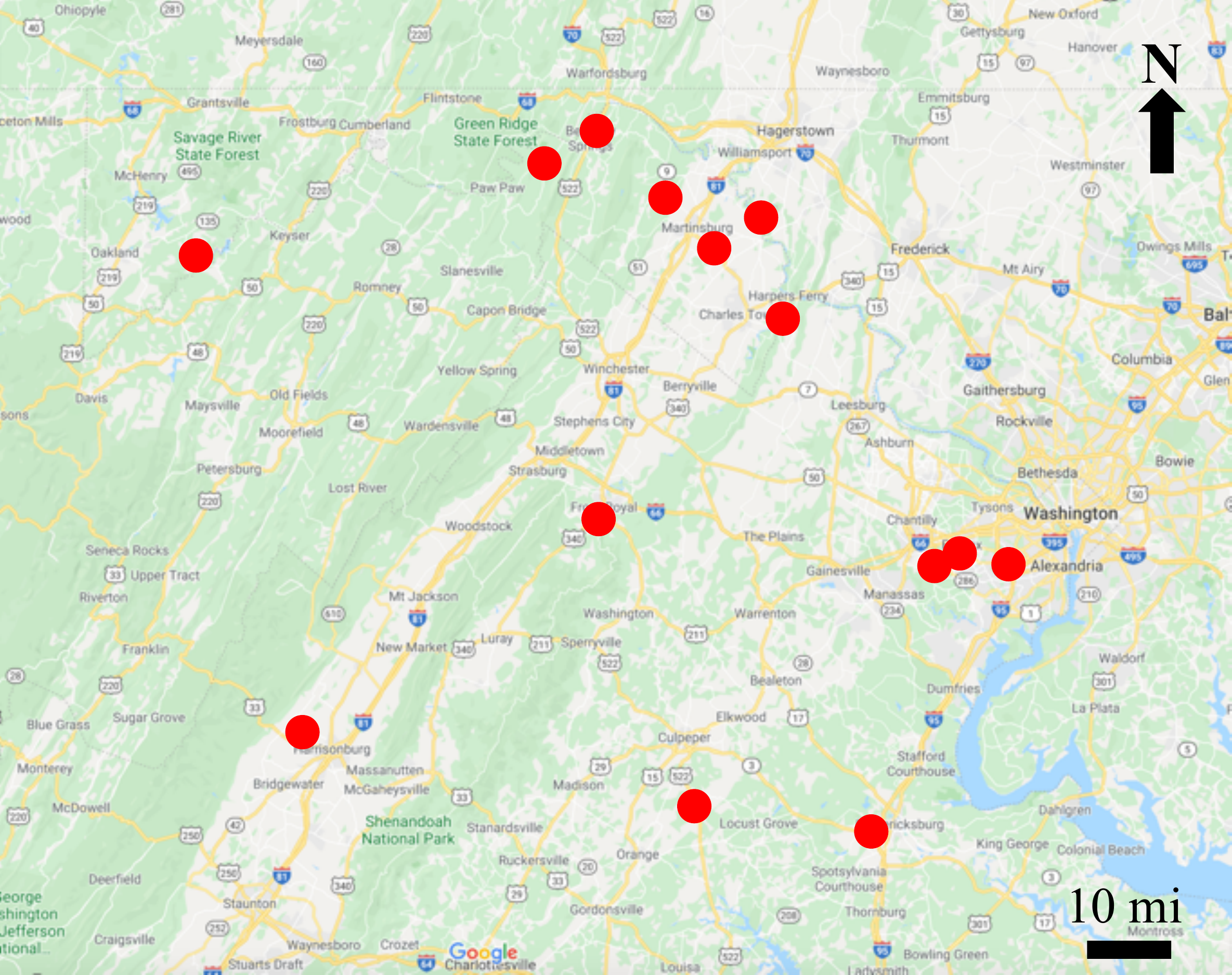}
	\caption{Location map of the streams in our study.}
	\label{fig:location_map1}
\end{figure}

In this section, we apply the proposed NVAR method to the case study of surface water quality data, focusing on observed total nitrogen (TN) loads from the U.S. Geological Survey (USGS) stream gauges.
The daily TN data (1990 to current) for the states of Virginia and West Virginia were downloaded.
We conducted an initial analysis of data availability and temporal consistency.
First we transfer the original data to a monthly data by using each month's maximum value as some months have more than one measurement.
Next we select a subset of data, which has $p$ time series, and does not contain any missing values over some $n$ consecutive months.
Among these selected possible datasets, we choose the one with the largest sample size, $pn$.
As a result, a total of $p=14$ monitoring sites were selected as the dataset for analysis, and the length of time series is $n=73$.
Figure \ref{fig:location_map1} reports the locations of the sites of the 14 streams of study.

We analyze this nitrogen dataset using the proposed NVAR method in comparison with the BVAR method and the LASSO method.
To evaluate the performance of the methods in comparison, we partition the data into the training data and test data.
For each time series, the beginning $80\%$ data are used as training data and the later $20\%$ data are used as test data.
The one-step ahead prediction is used to calculate the mean squared prediction errors for the methods in comparison.
For the NVAR method,
the definition of bandwidth $k_{NVAR}$ is a little different from the random structure.
The $k_{NVAR}$ equals the number of selected neighbors that are closest to a stream of interest.
Table~\ref{tab: appl_water} reports the performance of the methods in comparison.
It is seen that the proposed NVAR method performs better than the BVAR method in terms of mean squared prediction error,
and both the NVAR method and the BVAR method have much lower values of mean squared prediction error in comparison with that the LASSO method.

Note that the order of streams needs to be specified for applying the BVAR method, and the streams are ordered by their longitude for the BVAR method in Table~\ref{tab: appl_water}.
It means that, when using the BVAR method for analyzing the nitrogen data, its performance depends on how to specify the order of streams.
In contrast, the proposed NVAR method accommodates the natural distance measure to allow the analysis invariant on the order of the streams.
Table~\ref{tab: water_bvar_index} shows the results of the BVAR method using different directions to order the streams.
The PCA 1 index chooses the direction which explains the largest variety of the 2-D location (i.e., longitude and latitude).
The PCA 2 index uses the direction which is perpendicular to the PCA 1 direction.
As shown in the Table~\ref{tab: water_bvar_index}, the performance of the BVAR method varies among different directions.
It is seen that the order of streams has a significant effect on the performance of the BVAR method.
The prediction performance of the BVAR method is not as good as the proposed NVAR method.
Note that the NVAR method incorporates 2-D information instead of 1-D, thus it can include adequate information and is more robust than the BVAR method.

\begin{table}[H]
	\caption{\label{tab: appl_water}Mean square prediction error (MSPE) of NVAR, BVAR, LASSO for the stream data. A upper bound is set for the number of selected variables, which is $p/2$.}
	\centering
	\scriptsize
	\begin{tabular}{c c|m{0.6cm}|m{0.7cm}|m{0.7cm}|m{0.6cm}|m{0.7cm}|m{0.7cm}|m{0.7cm}|m{0.7cm}}
		\hline \hline
		& & \multicolumn{3}{c|}{NVAR} & \multicolumn{3}{c|}{BVAR} & \multicolumn{2}{c}{LASSO} \\
		\hline
		$p$ & $n$ & Est. band width  & MSPE & Comp. Time & Est. band width & MSPE & Comp. Time & MSPE & Comp. Time \\
		\hline
		14 & 73 & 7 & 0.746 & 0.092 & 3  & 1.026 & 0.053 & 4.787 & 0.411 \\
		\hline \hline
	\end{tabular}
\end{table}

\begin{table}[H]
	\caption{\label{tab: water_bvar_index}Mean square prediction error (MSPE) of the BVAR with different index for the stream data. A upper bound is set for the number of selected variables, which is $p/2$.}
	\centering
	\scriptsize
	\begin{tabular}{c c|m{1cm}|m{1cm}|m{1cm}|m{1cm}|m{1cm}|m{1cm}|m{1cm}|m{1cm}}
		\hline \hline
		& & \multicolumn{2}{m{2.3cm}|}{BVAR:\newline \textbf{ longitude index}} & \multicolumn{2}{m{2.2cm}|}{BVAR:\newline \textbf{latitude index}} & \multicolumn{2}{m{2cm}|}{BVAR:\newline \textbf{PCA 1 index}} & \multicolumn{2}{m{2cm}}{BVAR:\newline \textbf{PCA 2 index}} \\
		\hline
		$p$ & $n$ & Est. band width & MSPE & Est. band width & MSPE & Est. band width & MSPE & Est. band width & MSPE\\
		\hline
		14 & 73 & 3 & 1.026 & 3 & 0.813 & 3 & 1.069 & 3 & 0.941 \\
		\hline \hline
	\end{tabular}
\end{table}

\section{Discussion}
In this work, we have presented the neighborhood vector autoregression model, which utilizes the underlying distances among the time series based on the inherent setting of the problem.
We have generalized the model assumption in \citet{BandedVAR} by extending the notion of ``band'' to the notion of a ``neighborhood''.
The notion of neighborhood can be quite general under a distance or dissimilarity measure on where the data of multiple time series are collected.
With the aid of the Bayesian information criterion to choose an appropriate neighborhood size, the proposed NVAR method uses least squares for parameter estimation, thus can outperform penalization-based algorithms (e.g. \citet{Lozano2009,
	Bolstad2011}) in terms of computing efficiency.
We also investigate the theoretical properties of the proposed method under some regularity conditions.
Our theoretical studies show that the optimum neighborhood distance selected by the NVAR method converges to the true neighborhood distance, and the estimated coefficient matrix converges to the true coefficient matrix.
The simulation study and case study of stream nitrogen data show the NVAR method outperforms the BVAR method and the LASSO method, and the NVAR method is more robust than the BVAR method.
In particular, it is seen that the proposed method can gain prediction  accuracy by borrowing information from the neighborhood streams.

\subsection{Industrial applications}
This method can be directly applied to multiple problems in the industrial engineering literature and shows the potential for integrating sophisticated modeling of multivariate timeseries with an easily accessible idea for selecting relevant ``neighborhoods'' that will be especially useful in practical applications. In advanced manufacturing paradigms, the notion of ``neighborhood'' depends on the particular application, i.e., spatial distance, manufacturing process similarity, upstream and downstream flows, relational networks in the process etc. may all be considered in different applications, and our method is flexible enough to accommodate these different ideas. At the same time, the results of our method can be directly applied to solve specific management-related issues by proposing an ideal, but practical solution to the problem of dependencies in complex processes.

\subsection{Limitations}
A potential limitation of the proposed NVAR method is the assumption that the dependency (e.g., spatial dependency) among the time series must exist, and it can be captured by a distance or dissimilarity matrix.
When there is no such dependency among the time series or the distance matrix is unknown, then the NVAR method might not show advantages over other methods.

There are a few directions for future research.
First, we would like to study the NVAR method in terms of the estimation of auto-covariance matrix, and compare the performance with other methods in terms of estimation accuracy, efficiency, and theoretical convergence.
Second, our current approach adopts the BIC to choose the size of the neighborhood, and then the parameters are estimated by the ordinary least squares. Alternatively, we can use the penalized least squares for parameter estimation. Under this situation, it will be interesting to investigate what will be appropriate penalty functions for the neighbor vector autoregression model. It will also be interesting to examine the theoretical properties on the estimation accuracy when penalized estimation is involved.
Third, to address the limitation of relying on a distance matrix, one potential remedy is to combine covariance/precision matrix estimation with the NVAR method. It is reasonable to view the multivariate time series as a graph, then the conditional dependency can be viewed as the notion of ``neighborhood''.
Thus, the neighborhood of time series can be identified with some sparse covariance/precision matrix estimation method, which is a substitute when the distance matrix is unknown.

\section*{Data Availability Statement}
Data openly available in a public repository that does not issue DOIs. The data that support the findings of this study are openly available in National Water Information System (NWIS) at https://waterdata.usgs.gov/nwis.

\clearpage
\bibliographystyle{rss}
\bibliography{NVAR}

\clearpage
\begin{appendices}
	\section{Supplementary Material}
	The proofs of \cref{thm1} and \cref{thm2} follow a similar technical development as those in Guo. et al. (2016).
	The difference is that the bandwidth $k_0$ is replaced with neighborhood distance $d_0$, so the structure of coefficient matrix $A_s$ is different.
	Some regularity conditions also need to be changed.
	
	\subsection{Proof of Theorem 1}
	Without loss of generality, we consider the NVAR(1) model with $||A||_1 \le \delta < 1$. Our goal is to prove that pr($\hat{d}=d_0)\rightarrow 1$, i.e., pr($\hat{d}\ne d_0)\rightarrow 0$. If $\hat{d} \ne d_0$, then either $\hat{d}>d_0$ or $\hat{d}<d_0$ holds. Hence it suffices to show that pr($\hat{d}<d_0)\rightarrow 0$ and pr($\hat{d}>d_0)\rightarrow 0$. Our proof follows the arguments in Guo. et al. (2016).
	
	Consider the first case. Observe that pr($\hat{d}<d_0$) $\le$ pr($\hat{d}_i<d_0$) for some $i  \in \{1,...,p\}$ and the event ($\hat{d}_i<d_0$) imply $\{\min_{d<d_0}\text{BIC}(d,i)<\text{BIC}(d_0,i)\}$. To prove pr($\hat{d}\ne d_0)\rightarrow 0$, we only need to show that
	\begin{align*}
	\text{pr}\{\min_{d<d_0}\text{BIC}(d,i)<\text{BIC}(d_0,i)\} \rightarrow 0
	\end{align*}
	for some $i$. Suppose that we have shown that there exists a constant $\eta > 0$ and an event $\mathcal{A}_n$ such that pr$(\mathcal{A}_n)\rightarrow 1$ as $n\rightarrow \infty$ and on the event $\mathcal{A}_n$,
	\begin{align}
	\label{A1}
	\text{RSS}(d,i)-\text{RSS}(d_0,i)\ge \eta \text{RSS}(d_0,i)\sum_{j \in \mathcal{N}_{i}^{d_0}}(a_{i,j}^2)
	\end{align}
	for sufficiently large $n$, where $a_{j,k}$ is the $(j,k)$-element of $A_1$. On the event $\mathcal{A}_n$ with large $n$, $\text{logRSS}(d,i)-\text{logRSS}(d_0,i)\ge \text{log}\{1+\eta \sum_{j \in \mathcal{N}_{i}^{d_0}}(a_{i,j}^2)\}$. Note that $\text{log}(1+x)\ge \text{min}(0.5x,\text{log}2)$ for any $x>0$. consequently, with probability tending to one, $\text{logRSS}(d,i)-\text{log RSS}(d_0,i)$ can be further bounde below by $\text{min}(0.5\eta \sum_{j \in \mathcal{N}_{i}^{d_0}}a_{i,j}^2,\text{log}2)$. Condition 2 implies that for some $i^* \in\{1,...,p\}$, $0.5\eta\sum_{j \in \mathcal{N}_{i^*}^{d_0}}a_{i^*,j}^2 \gg C_n \tau_{i^*}^{d_0} n^{-1} \text{log}(p \vee n)$ as $n\rightarrow \infty$, where $\tau_{i^*}^{d_0}=|\mathcal{N}_{i^*}^{d_0}|$. Hence, it follows that, with probability tending to 1,
	\begin{align*}
	\min_{d<d_0}\text{BIC}(d,i^*)-\text{BIC}(d_0,i^*)&=\text{logRSS}(d,i^*)-\text{logRSS}(d_0,i^*)+C_n (\tau_{i^*}^{d}-\tau_{i^*}^{d_0}) n^{-1} \text{log}(p \vee n)\\
	&>\text{min}(0.5\eta\sum_{j \in \mathcal{N}_{i^*}^{d_0}}a_{i^*,j}^2,\text{log}2)-C_n \tau_{i^*}^{d_0} n^{-1} \text{log}(p \vee n)\\
	&\gg 0.
	\end{align*}
	where $p\vee n=\mathrm{max}(p,n)$, $\mathcal{N}_{i^*}^{d}$ is the length of non-zero elements in the $i^*$-th row of $A_1$ with $d$-neighborhood, $\mathcal{N}_{i^*}^{d_0}$ is the length of non-zero elements in the $i^*$-th row of $A_1$ with $d_0$-neighborhood. Hence, $\text{pr}\{\min_{d<d_0}\text{BIC}(d,i)<\text{BIC}(d_0,i)\}\rightarrow 0$ and thus pr($\hat{d}<d_0)\rightarrow 0$.\\
	
	Let us prove \cref{A1}. For $d<d_0$, denote $H_{i,d}=X_{i,d}(X_{i,d}^TX_{i,d})^{-1}X_{i,d}^T$, $X_{i,d_0}=(S_{i,d},X_{i,d})$ and $\beta_{i,d_0}=(b_i^T,\beta_{i,d}^T)^T$, where $X_{i,d}$, $\beta_{i,d}$, $S_{i,d}$ are defined as below
	\begin{align*}
	&y_i(t)=\sum_{j \in \mathcal{N}_{i}^{d_0}} A_1(i,j)y_j(t-1)+e_i(t),\\
	&y_i=X_{i,d}\beta_{i,d}+e_i,\\
	&\text{Let }\{y_j(t-1)\}|_{j \in \mathcal{N}_{i}^{d}}\text{ be a column vector},\\
	&\text{then }X_{i,d}=\Big\{\{y_j(n-1)\}|_{j \in \mathcal{N}_{i}^{d}},\{y_j(n-2)\}|_{j \in \mathcal{N}_{i}^{d}},...,\{y_j(1)\}|_{j \in \mathcal{N}_{i}^{d}}\Big\}^T,\\
	&\beta_{i,d}=\Big\{A_1(i,j)\Big\}\Big|_{j \in \mathcal{N}_{i}^{d}},\\
	&S_{i,d}=\Big\{\{y_j(n-1)\}|_{j \in \mathcal{N}_{i}^{d_0}\backslash \mathcal{N}_{i}^{d}},\{y_j(n-2)\}|_{j \in \mathcal{N}_{i}^{d_0}\backslash \mathcal{N}_{i}^{d}},...,\{y_j(1)\}|_{j \in \mathcal{N}_{i}^{d_0}\backslash \mathcal{N}_{i}^{d}}\Big\}^T,\\
	&\{j \in \mathcal{N}_{i}^{d_0}\backslash \mathcal{N}_{i}^{d}\} = \{j \in \mathcal{N}_{i}^{d_0} \text{ and } j \notin \mathcal{N}_{i}^{d}\}.
	\end{align*}
	Then $\text{RSS}(d,i)=y_i^T(I_{n-1}-H_{i,d})y_i$, and by \cref{lma1}(ii) or \cref{lma2}(ii), we have
	\begin{align*}
	\text{RSS}(d,i)-\text{RSS}(d_0,i)=b_i^TS_{i,d}^T(I_{n-1}-H_{i,d})S_{i,d}b_i + o_P(1).
	\end{align*}
	From \cref{lma1}(ii) or \cref{lma2}(ii) and \cref{lma3}, there exists a small constant $\eta>0$ such that, with probability tending to one,
	\begin{align*}
	\lambda_{min}\{S_{i,d}^T(I_{n-1}-H_{i,d})S_{i,d}\}>\eta(1+\eta)n\sigma_i^2,
	\end{align*}
	and $\text{RSS}(d_0,i) \le (1+\eta)n\sigma_i^2$. Therefore, \cref{A1} follows.\\
	
	Now let us prove the second case, pr($\hat{d}>d_0)\rightarrow 0$. For $d>d_0$, set
	\begin{align*}
	X_{i,d}=(S_{i,d},X_{i,d_0}),\ \beta_{i,d}=(0^T,\beta_{i,d_0}^T)^T, \text{ and } \tilde{S}_{i,d}=(I_{n-1}-H_{i,d_0}S_{i,d}).
	\end{align*}
	Let $\eta$ be an arbitrary but fixed positive constant and define
	\begin{align*}
	\mathcal{B}_n=\Big\{\inf_{d_0\le d\le d_{max}} \inf_{1\le i \le p} \frac{\text{RSS}_i(k)}{n\sigma_i^2}>(1-\eta)\Big\},
	\end{align*}
	\begin{align*}
	\mathcal{C}_n=\bigcup_{1 \le i \le p,\ d_0 \le d \le d_{max}}\Big\{\lambda_{\text{min}}^{-1}(n^{-1}\tilde{S}_{i,d}^T\tilde{S}_{i,d})<\kappa_1^{-1}(1+\eta), \sup_{1\le j \le d-d_0}\big|(n^{-1}S_{i,d}^TS_{i,d})_{jj}\big|<\kappa_2(1+\eta)\Big\}.
	\end{align*}
	We first give an upper bound on $\text{RSS}(d_0,i)-\text{RSS}(d,i)$ for $d>d_0$. For each $i$, $\text{RSS}(d,i)$ can be rewritten as
	\begin{align*}
	\text{RSS}(d,i)=\inf_{b} ||y_i-X_{i,d}b||^2=\inf_{b_1,b_2} ||y_i-X_{i,d}b_1-S_{i,d}b_2||^2.
	\end{align*}
	where $b$ is the estimator for $\beta_{i,d}$. It can be verified that $\text{RSS}(d_0,i)=||(I_{n-1}-H_{i,d_0})y_i||^2$ and $\text{RSS}(d,i)=\text{RSS}(d_0,i)-||\tilde{S}_i^{(d)}\hat{b}_2||^2$, where $\hat{b}_2=\Big(\tilde{S}_{i,d}^T\tilde{S}_{i,d}\Big)^{-1}\tilde{S}_{i,d}^Te_i$, and $e_i$ is the residual for $i$-th time series. Then on the event $\mathcal{C}_n$ we have
	\begin{align*}
	\text{RSS}(d_0,i)-\text{RSS}(d,i)&=e_i^T\tilde{S}_{i,d}(\tilde{S}_{i,d}^T\tilde{S}_{i,d})^{-1}\tilde{S}_{i,d}^Te_i\\
	&\le \kappa_1^{-1}(1+\eta)|\tau_{i}^{d}-\tau_{i}^{d_0}|\ \sup_{j,d\le p} |n^{-1/2}e_j^T(I_{n-1}-H_{i,d_0})x_{(d)}|^2.
	\end{align*}
	Define
	\begin{align*}
	\mathcal{D}_n=\Big\{\sup_{j,d\le p} \big|n^{-1/2}e_j^T(I_{n-1}-H_{i,d_0})x_{(d)}\big|^2\sigma_i^{-2}<\frac{\kappa_1(1-\eta)}{1+\eta}C_n\text{log}(p\vee n)\Big\}.
	\end{align*}
	On the set $\mathcal{B}_n \cap \mathcal{C}_n \cap \mathcal{D}_n$, for all $d$ with $d_0\le d \le d_{max}$,
	\begin{align*}
	\text{RSS}(d_0,i)-\text{RSS}(d,i)&< \sigma_i^2(1-\eta)|\tau_{i}^{d}-\tau_{i}^{d_0}|C_n \text{log}(p\vee n)\\
	&<\text{RSS}(d,i)C_n|\tau_{i}^{d}-\tau_{i}^{d_0}|n^{-1}\text{log}(p\vee n).
	\end{align*}
	Note that $\text{log}(1+x)\le x$ for any $x>0$. Hence, for all $d$ with $d_0\le d \le d_{max}$, on the set $\mathcal{B}_n \cap \mathcal{C}_n \cap \mathcal{D}_n$,
	\begin{align*}
	\text{BIC}(d,i)-\text{BIC}(d_0,i)&= \text{log RSS}(d,i) - \text{log RSS}(d_0,i) + C_n|d^m(i)-d_0^m(i)|n^{-1}\text{log}(p\vee n)\\
	&\ge -\text{log}\Big(1+C_n|\tau_{i}^{d}-\tau_{i}^{d_0}|n^{-1}\text{log}(p\vee n)\Big)\\
	&\ \ \ \ +C_n|\tau_{i}^{d}-\tau_{i}^{d_0}|n^{-1}\text{log}(p\vee n)
	\end{align*}
	which indicates that over the set $\mathcal{B}_n \cap \mathcal{C}_n \cap \mathcal{D}_n$, we have that $\hat{d}\le d_0$. To prove that $\text{pr}(\hat{d}>d_0)\rightarrow 0$, it is sufficient to show that $\text{pr}\big\{(\mathcal{B}_n \cap \mathcal{C}_n \cap \mathcal{D}_n)^c\big\}\rightarrow 0$. In fact, it follows from \cref{lma3} and \cref{lma1} or \cref{lma2}(i), that $\text{pr}(\mathcal{B}_n^c)\rightarrow 0$ and $\text{pr}(\mathcal{C}_n^c)\rightarrow 0$. It remains to show that $\text{pr}(\mathcal{D}_n^c)\rightarrow 0$. Let $\hat{\Sigma}_{i,d}=n^{-1}X_{i,d}^TX_{i,d}$, $\Sigma_{i,d}=n^{-1}E\big(X_{i,d}^TX_{i,d}\big)$, where $E(X)$ denotes the expectation of $X$. Set $\tilde{H}_{i,d}=n^{-1}X_{i,d}\Sigma_{i,d}^{-1}X_{i,d}^T$, and $\tilde{x}_{(d)}=(I_{n-1}-\tilde{H}_{i,d})x_{(d)}$. On the event $\mathcal{D}_n$, we obtain that
	\begin{align*}
	\sup_{j,d\le p} |e_j^T(I_{n-1}-H_{i,d_0})x_{(d)}| &\le \sup_{j,d\le p} |e_j^T\tilde{x}_{(d)}| + \sup_{j,d\le p} |e_j^T(I_{n-1}-\tilde{H}_{i,d_0})x_{(d)}|\\
	&\le \sup_{j,d\le p} |e_j^T\tilde{x}_{(d)}| \\
	&\ \ + \sup_{j,d\le p} ||e_j^TX_{i,d_0}||_2 \ ||\Sigma_{i,d_0}^{-1}||_2 \ ||\hat{\Sigma}_{i,d_0}^{-1}||_2 \ ||\hat{\Sigma}_{i,d_0}-\Sigma_{i,d_0}||_2 \ ||X_{i,d_0}^Tx_{(d)}||_2 \\
	&\le \sup_{j,d\le p} |e_j^T\tilde{x}_{(d)}| + d_0\kappa_1^{-2} \kappa_2 (1+\eta)^2 \sup_{j,d\le p} \big|e_j^Tx_{(d)}\big|\cdot ||\hat{\Sigma}_{i,d_0}-\Sigma_{i,d_0}||_2,
	\end{align*}
	where $\sup_{1\le d \le p}(n^{-1}x_{(d)}x_{(d)}^T)\le \kappa_2 (1+\eta)$ is used in the above inequality. Hence, it follows from \cref{lma1,lma2}, together with Condition 3, that $\text{pr} (\mathcal{D}_n^c)\rightarrow 0$ as $n\rightarrow \infty$. Then pr($\hat{d}>d_0)\rightarrow 0$. $\square$

	 \vspace{-.5cm}
	\subsection{Proof of Theorem 2}
	Without loss of generality, we consider the case of order 1, i.e. NVAR(1) only. It is shown in the Theorem 1 that $\text{pr}(\hat{d}=d_0)\rightarrow 1$ as $n\rightarrow \infty$. Thus it is sufficient to consider the set $\mathcal{A}_n=\{\hat{d}=d_0\}$. Over the set $\mathcal{A}_n$, for each $i$,
	\begin{align}
	\hat{\beta}_i-\beta_i=(X_i^TX_i)^{-1}x_i^Te_i
	\end{align}
	For each $i$, the law of large numbers for the stationary process case yields that $n^{-1}X_i^TX_i$ converges to a positive matrix almost surely, and furthermore, with probability tending to one, $\lambda_{min}(n^{-1}X_i^TX_i)$ is bounded away from zero. As a matter of fact, if we define
	\begin{align*}
	\mathcal{B}_n=\bigcap_{1\le i \le p} \Big\{\lambda_{min}(n^{-1}X_i^TX_i)>\kappa_1(1-\eta)\Big\}
	\end{align*}
	with a small constant $\eta \in (0,1)$, then it follows from by \cref{lma1,lma2} under different moment conditions that $P\{\mathcal{B}_n\}\rightarrow 1$ as $n\rightarrow \infty$. Hence, over the event $\mathcal{A}_n \cup \mathcal{B}_n$,
	\begin{align*}
	\Big|\Big|\hat{\beta}_i-\beta_i\Big|\Big|_2^2 &\le \kappa_1^{-2} (1-\eta)^{-2}n^{-2} ||e_i^TX_i||_2^2,\\
	&=C_1n^{-2} ||e_i^TX_i||_2^2,
	\end{align*}
	where $C_1=\kappa_1^{-2} (1-\eta)^{-2}>0$. It is not hard to see from \cref{lma1}(ii) or \cref{lma2}(ii) that, for all $1\le i \le p,\ n^{-1}E||X_i^Te_i||_2^2\le C_2$ with some constant $C_2>0$. Therefore, for a large positive constant $C$, we obtain that
	\begin{align*}
	\text{pr}\Bigg(\bigg|\bigg|\hat{A}_1-A_1\bigg|\bigg|_F^2>Cn^{-1}p\Bigg)&= \text{pr}\Bigg(\bigg|\bigg|\hat{A}_1-A_1\bigg|\bigg|_F^2>Cn^{-1}p, \mathcal{A}_n \cup \mathcal{B}_n\Bigg)\\
	&\ \ \ \ +\text{pr}\Bigg(\bigg|\bigg|\hat{A}_1-A_1\bigg|\bigg|_F^2>Cn^{-1}p, (\mathcal{A}_n \cup \mathcal{B}_n)^c\Bigg)\\
	&\le (Cp)^{-1}n(C_1n^{-2})E\Bigg(\sum_{i=1}^{p} ||X_i^Te_i||_2^2\Bigg)\\
	&\ \ \ \ +\text{pr}\Bigg(\bigg|\bigg|\hat{A}_1-A_1\bigg|\bigg|_F^2>Cn^{-1}p, (\mathcal{A}_n \cup \mathcal{B}_n)^c\Bigg)\\
	&\le C_1C_2C^{-1} +o(1)
	\end{align*}
	For a sufficiently large C, we have $\text{pr}\big(||\hat{A}_1-A_1||_F^2>Cn^{-1}p\big)\rightarrow 0$. Thus  the convergence rate of $||\hat{A}_1-A_1||_F$ is established.
	
	Now Let us derive the convergence rate of $||\hat{A}_1-A_1||_2$. For any matrix $B$, $||B||_2^2 \le ||B||_1 ||B||_{\infty}$. Hence, on the event $\mathcal{A}_n$,
	\begin{align*}
	||\hat{A}_1-A_1||_2 &\le \sqrt{||\hat{A}_1-A_1||_1} \sqrt{||\hat{A}_1-A_1||_{\infty}}\\
	&\le \tau_{i}^{d_0} \sup_{i,j\le p} |\hat{\beta}_{ij}-\beta_{ij}|,
	\end{align*}
	where $\hat{\beta}_{ij}$ and $\beta_{ij}$ are the $j$-th element of $\hat{\beta}_{i}$ and $\beta_{i}$, respectively. Observe from (3) that
	\begin{align*}
	\sup_{i,j\le p} |\hat{\beta}_{ij}-\beta_{ij}|=\kappa_1^{-1}(1-\eta)^{-1} \tau_{i}^{d_0} n^{-1} \big(\sup_{i,j\le p}|e_i^Tx_{(j)}|\big),i=1,...,p.
	\end{align*}
	Hence, using \cref{lma1}(ii) or \cref{lma2}(ii), we have
	\begin{align*}
	\sup_{i,j\le p} |\hat{\beta}_{ij}-\beta_{ij}|=O_P\Big\{\big(n^{-1}\text{log}p\big)^{1/2}\Big\},
	\end{align*}
	which shows that
	\begin{align*}
	||\hat{A}_1-A_1||_2=O_P\Big\{\big(n^{-1}\text{log}p\big)^{1/2}\Big\}.
	\end{align*}
	Then the proof is done. $\square$
	
	\section{Technical lemmas}
	The proof of \cref{lma1,lma2,lma3} can be found in Guo. et al. (2016), which are corresponding to Lemma 5 to 7. In addition, the regularity conditions should be replaced with our regularity conditions, and the lemmas still hold.
	
	\begin{lemma}
		\label{lma1}
		Suppose that Conditions (1)-(3) and 4(i) hold.\\
		(i) For $j,d=1,\dots,p$, there exist positive constants $C_1$, $C_2$, and $C_3$ free of $(j,d,n,p)$ such that
		\begin{align*}
		\text{pr}\bigg(\bigg|\hat{\Sigma}_{jd}-\Sigma_{jd}\bigg|>x\bigg) \le \frac{C_1n}{(nx)^q}+C_2\text{exp}(-C_3nx^2)
		\end{align*}
		holds for $x>0$; consequently, this leads to the following uniform convergence rate:
		\begin{align*}
		\sup_{1\le j,d\le p}\bigg|\hat{\Sigma}_{jd}-\Sigma_{jd}|\bigg| = O_P\bigg\{(n^{-1}\text{log}p)^{1/2}\bigg\}.
		\end{align*}
		(ii) For $j,d=1,\dots,p$, there exist positive constants $C_1$, $C_2$, and $C_3$ free of $(j,d,n,p)$ such that
		\begin{align*}
		\text{pr}\bigg(\bigg|e_j^Tx_{(k)}\bigg|>x\bigg) \le \frac{C_1n}{x^{2q}}+C_2\text{exp}(-C_3x^2)
		\end{align*}
		holds for $x>0$; in particular, we have:
		\begin{align*}
		\sup_{1\le j,d\le p}\bigg|e_j^Tx_{(k)}\bigg| = O_P\bigg\{(n\text{log}p)^{1/2}\bigg\}.
		\end{align*}
	\end{lemma}
	
	\begin{lemma}
		\label{lma2}
		Suppose that Conditions (1)-(3) and 4(ii) hold.\\
		(i) \begin{align*}
		\sup_{1\le j,d\le p}\bigg|\hat{\Sigma}_{jd}-\Sigma_{jd}|\bigg| = O_P\bigg\{(n^{-1}\text{log}p)^{1/2}\bigg\}.
		\end{align*}
		(ii) \begin{align*}
		\sup_{1\le j,d\le p}\bigg|e_j^Tx_{(k)}\bigg| = O_P\bigg\{(n\text{log}p)^{1/2}\bigg\}.
		\end{align*}
	\end{lemma}
	
	\begin{lemma}
		\label{lma3}
		Suppose that Conditions (1)-(3) and 4(i) or 4(ii) hold. Then for each finite $d$ with $d \ge d_0$,\\
		\begin{align*}
		\sup_{1\le i \le p}\bigg(\bigg|\frac{\text{RSS}(d,i)}{n\sigma_i^2}-1\bigg| = O_P\bigg\{(n^{-1}\text{log}p)^{1/2}\bigg\}.
		\end{align*}
		as $n\rightarrow \infty$, where $\text{RSS}(d,i)$ is defined in the main article and $\sigma_i^2$ is the $(i,i)$-th element of $\Sigma_e$.
	\end{lemma}

\end{appendices}

\end{document}